\documentclass[trackchanges]{aastex7}

\usepackage{txfonts}
\usepackage{hyperref}

\begin{document}


\title{The contribution of the color space in LSST-like photometry for the selection of extragalactic globular cluster candidates}

\author[orcid=0009-0001-0407-8134,sname=Schweder-Souza]{Nicholas Schweder-Souza}
\altaffiliation{}
\affiliation{Departamento Acadêmico de Física, Universidade Tecnológica Federal do Paraná, Av. Sete de Setembro 3165, Curitiba, Brazil}
\affiliation{Laboratório Interinstitucional de e-Astronomia (LIneA), Rua Gal.
José Cristino 77, Rio de Janeiro, RJ, 20921-400, Brazil} 
\email[show]{nicholassouza@alunos.utfpr.edu.br}  

\author[orcid=0000-0003-3220-0165,sname=Chies-Santos]{Ana L. Chies-Santos} 
\altaffiliation{}
\affiliation{Instituto de Física, Universidade Federal do Rio Grande do Sul, Av. Bento Gonçalves 9500, Porto Alegre, RS 90040-060, Brazil}
\affiliation{Laboratório Interinstitucional de e-Astronomia (LIneA), Rua Gal.
José Cristino 77, Rio de Janeiro, RJ, 20921-400, Brazil} 
\email{ana.chies@ufrg.br}

\author[orcid=0000-0001-7207-4584,sname=Souza]{Rafael S. de Souza}
\affiliation{Centre for Astrophysics Research, University of Hertfordshire, College Lane, Hatfield, AL10~9AB, UK}
\affiliation{Instituto de Física, Universidade Federal do Rio Grande do Sul, Av. Bento Gonçalves 9500, Porto Alegre, RS 90040-060, Brazil}
\affiliation{Department of Physics \& Astronomy, University of North Carolina at Chapel Hill, NC 27599-3255, USA}
\email{rd23aag@herts.ac.uk}

\author[orcid=0000-0002-8532-4025,sname=Dage]{Kristen C. Dage}
\altaffiliation{}
\affiliation{International Centre for Radio Astronomy Research, Curtin University, GPO Box U1987, Perth, WA 6845, Australia}
\email{kristen.dage@curtin.edu.au}

\author[orcid=0000-0002-4102-1751,sname=Bonatto]{Charles J. Bonatto}
\altaffiliation{}
\affiliation{Instituto de Física, Universidade Federal do Rio Grande do Sul, Av. Bento Gonçalves 9500, Porto Alegre, RS 90040-060, Brazil}
\affiliation{Laboratório Interinstitucional de e-Astronomia (LIneA), Rua Gal.
José Cristino 77, Rio de Janeiro, RJ, 20921-400, Brazil}
\email{charles.bonatto@ufrgs.br}

\author[orcid=0000-0003-0812-9928,sname=Caso]{Juan P. Caso}
\altaffiliation{}
\affiliation{Facultad de Ciencias Astron\'omicas y Geof\'isicas de la Universidad Nacional de La Plata, and Instituto de Astrof\'isica de La Plata, Paseo del Bosque S/N, B1900FWA La Plata, Argentina}
\affiliation{Consejo Nacional de Investigaciones Cient\'ificas y T\'ecnicas, Godoy Cruz 2290, C1425FQB, Ciudad Aut\'onoma de Buenos Aires, Argentina}
\email{jpceda@gmail.com}

\author[orcid=0000-0003-2072-384X,sname=Cantiello]{Michele Cantiello}
\altaffiliation{}
\affiliation{INAF Osservatorio Astronomico d'Abruzzo, Via Maggini, 64100 Teramo, Italy}
\email{michele.cantiello@inaf.it}

\author[orcid=0009-0005-7299-4168,sname=Santos-Lopes]{Pedro dos Santos-Lopes}
\altaffiliation{}
\affiliation{Instituto de Física, Universidade Federal do Rio Grande do Sul, Av. Bento Gonçalves 9500, Porto Alegre, RS 90040-060, Brazil}
\affiliation{Laboratório Interinstitucional de e-Astronomia (LIneA), Rua Gal.
José Cristino 77, Rio de Janeiro, RJ, 20921-400, Brazil} 
\email{00325245@ufrgs.br}

\author[orcid=0009-0008-4034-7670,sname=Floriano]{Pedro Floriano}
\altaffiliation{}
\affiliation{Instituto de Física, Universidade Federal do Rio Grande do Sul, Av. Bento Gonçalves 9500, Porto Alegre, RS 90040-060, Brazil}
\affiliation{Laboratório Interinstitucional de e-Astronomia (LIneA), Rua Gal.
José Cristino 77, Rio de Janeiro, RJ, 20921-400, Brazil}
\email{00318264@ufrgs.br}

\author[orcid=0000-0002-8139-7278, sname=Pacheco]{Thayse A. Pacheco}
\affiliation{Instituto de Física, Universidade Federal do Rio Grande do Sul, Av. Bento Gonçalves 9500, Porto Alegre, RS 90040-060, Brazil}
\affiliation{Universit\'e de Strasbourg, CNRS, Observatoire astronomique de Strasbourg, UMR 7550, 67000 Strasbourg, France}
\email{thayse.pacheco@ufrgs.br}

\author[orcid=0000-0001-8283-4591,sname=Rhode]{Katherine L. Rhode}
\altaffiliation{}
\affiliation{Department of Astronomy, Indiana University, Bloomington, IN 47405, USA}
\email{krhode@iu.edu}

\author[orcid=0000-0003-2767-0090,sname=Barmby]{Pauline Barmby}
\altaffiliation{}
\affiliation{Department of Physics \& Astronomy, The University of Western Ontario, London, ON N6A 3K7, Canada}
\email{pbarmby@uwo.ca}

\author[orcid=0000-0002-4989-0353,sname=Sobeck]{Jennifer Sobeck}
\altaffiliation{}
\affiliation{IPAC, Caltech, 1200 E. California Boulevard, Pasadena, CA 91125, USA}
\email{jsobeck@ipac.caltech.edu}

\author[orcid=0000-0001-8411-8783,sname=Ennis]{Ana I. Ennis}
\altaffiliation{}
\affiliation{Waterloo Centre for Astrophysics, University of Waterloo, Waterloo, Ontario, N2L 3G1, Canada}
\affiliation{Perimeter Institute for Theoretical Physics, Waterloo, Ontario N2L 2Y5, Canada}
\email{ana.ennis@uwaterloo.ca}

\author[orcid=0000-0001-7966-7606,sname=Ordenes-Brice\~no]{Yasna Ordenes-Brice\~no}
\altaffiliation{}
\affiliation{Instituto de Estudios Astrof\'isicos, Facultad de Ingenier\'ia y Ciencias, Universidad Diego Portales, Av. Ej\'ercito Libertador 441, Santiago, Chile} 
\email{yasna.ordenes@mail.udp.cl}

\author[orcid=0000-0002-9554-7660,sname=Saifollahi]{Teymoor Saifollahi}
\altaffiliation{}
\affiliation{Universit\'e de Strasbourg, CNRS, Observatoire astronomique de Strasbourg, UMR 7550, 67000 Strasbourg, France}
\affiliation{Centre national d'études spatiales (CNES), 2, Place Maurice Quentin, 75039, Paris, France}
\email{teymoor.saifollahi@astro.unistra.fr}

\author[orcid=0000-0003-3023-8362,sname=Gschwend]{Julia Gschwend}
\altaffiliation{}
\affiliation{Laboratório Interinstitucional de e-Astronomia (LIneA), Rua Gal.
José Cristino 77, Rio de Janeiro, RJ, 20921-400, Brazil} 
\email{julia@linea.org.br}

\author[orcid=0000-0003-3372-3638]{Niranjana P.}
\altaffiliation{}
\affiliation{Instituto de Física, Universidade Federal do Rio Grande do Sul, Av. Bento Gonçalves 9500, Porto Alegre, RS 90040-060, Brazil}
\affiliation{Laboratório Interinstitucional de e-Astronomia (LIneA), Rua Gal.
José Cristino 77, Rio de Janeiro, RJ, 20921-400, Brazil} 
\email{niranjana.prashantha@ufrgs.br}

\author[orcid=0000-0001-7319-297X,sname=Machado]{Rubens E. G. Machado}
\altaffiliation{}
\affiliation{Departamento Acadêmico de Física, Universidade Tecnológica Federal do Paraná, Av. Sete de Setembro 3165, Curitiba, Brazil} 
\email{rubensmachado@utfpr.edu.br}

\collaboration{all}{for the LSST Star Clusters working group}

\begin{abstract}
Globular clusters (GCs) are excellent tracers of their host galaxies' evolutionary histories. Traditional methods for identifying GCs in galaxies rely on cuts over photometric catalogs and can yield source lists with high levels of contamination from compact background galaxies and foreground stars. In an era when large-scale sky surveys produce photometry for millions of sources, it is essential to employ flexible and scalable tools to reliably identify GCs in external galaxies. To prepare for surveys like Rubin/LSST, we need to explore practical methodological improvements and quantify the limitations inherent in the datasets. This paper investigates the selection of point-like extragalactic GCs exclusively in the $ugrizY$ color space. We use archival data to assemble an LSST-like photometric catalog for the Fornax Cluster containing labeled confirmed GCs, galaxies, and stars. From this catalog, using principal component analysis and non-linear auto-encoders (AEs), we construct inputs to random forest and multi-layer perceptron classifiers. We show that selecting GCs using all the 15 available colors can lead to a minimum contamination rate of $\sim 30\%$, whereas the use of color-color diagrams may double such rate. If only the first 4 principal components of the colors are used instead, the same minimum contamination rate is achieved without increasing incompleteness. The AEs did not improve GC identification. To further reduce contamination and extract the full potential of LSST for star cluster studies, we argue for the need to augment photometric information with ancillary data (morphology from space-based missions and near-infrared photometry) before attempting to leverage more complex models.
\end{abstract}



\section{Introduction} \label{sec:intro}
A globular cluster (GC) is a very dense set of thousands to millions of stars, typically spanning masses from $10^4$ to $10^6$ $\rm M_{\odot}$ \citep{harris2014, baumgardt2018}. 
GCs found in local galaxies are typically very old, with ages comparable to a Hubble time \citep{usher2019, fahrion2020}.
They can be found in a wide variety of galaxy morphologies and carry rich information about the distant past of their hosts, thus allowing the study of the evolution of the systems \citep{beasley2020book}.
They have been observed to have half-light radii within the range $r_h \sim 2-5$ \citep{masters2010, brodie2011, caso2014}, which means that for the vast majority of telescopes, GCs rapidly become point-like sources the more distant the observed galaxy is.    

The correct identification of extragalactic GCs, as well as their incorporation into simulations, allows for a wide variety of scientific applications. It is possible to place constraints on the amount and distribution of baryonic and dark matter from the masses and spatial distributions of the GC systems of particular galaxies or galaxy clusters. When such results are interpreted in light of the properties of the host galaxies, pieces of the mass assembly history of these systems are revealed \citep{harris2013, hudson2014, forbes2018, burkert2020, valenzuela2021, saifollahi2022, zaritsky2022, reina-campos2023, diego2023, canossa2024, mirabile2024, dornan2025}. Also, distance estimates can be derived via the GC luminosity function (GCLF) if a universal form is assumed \citep{rejkuba2012}. In a complementary manner, kinematics and stellar population properties of extragalactic GCs are studied with spectroscopy and/or multi-band photometry to further detail the evolutionary paths of their host galaxies \citep{chies-santos2011b, annibali2018, usher2019, forbes2022, chies-santos2022, adamo2023, lomeli2024, grasser2024, usher2024}. 
Beyond that, GCs are also known to host unique high-energy, transient phenomena, from ultraluminous X-ray sources \citep{Maccarone07, Dage20} to fast radio bursts \citep{Kirsten22}.
As the study of GC systems has implications for a broad range of science cases, it is imperative to develop robust methodologies to identify extragalactic GC candidates.

Several systematic surveys that investigate extragalactic GC systems (among other objects) have been carried out in recent decades: e.g., the ACS Virgo Cluster Survey \citep{cote2004}; the ACS Fornax Cluster Survey \citep{jordan2007}; the optical-NIR survey of GCs in early-type galaxies by \cite{chies-santos2011a}; the Next Generation Virgo Cluster Survey \citep{ferrarese2012}; the SAGES Legacy Unifying Globulars and GalaxieS Survey \citep[SLUGGS;][]{brodie2014}; the Fornax Deep Survey \citep{cantielloFDS2021}.
Although different surveys develop distinct procedures to extract samples of extragalactic GC candidates, each with its specificities, there is a common ground, and some basic steps are well established. 
Since GCs can appear within the innermost parts of a galaxy out to many effective radii, a typical first step is to remove the diffuse stellar emission from the host galaxy. This is done either by fitting and subtracting a model of the galaxy light distribution or by smoothing the image and subtracting it from the original. A source detection algorithm (e.g., the one available in \texttt{SExtractor}) is then used to find objects that have fluxes above a specified threshold compared to the background noise.
Photometry is then performed, and a catalog of sources is produced, including their magnitudes and other quantities measured in the available filters.

The full catalog of the studied region contains, {\it a priori}, a heterogeneous population of unlabeled sources, including foreground stars, galaxies spanning a wide range of redshifts, and GCs. The methodology to identify the sources of different nature varies among different authors in the literature. The traditional approach is to apply linear cuts in the spaces of spectral energy distributions (SEDs), colors, and morphometric quantities. That is, lines and polygons are traced on color-color, color-magnitude diagrams (and other projections using, for example, FWHM, ellipticity) to define the regions associated with GCs and label candidates. This is the case for all the surveys previously cited, as well as other studies, including more recent works \citep{hargis2012, dabrusco2016, berkheimer2025, lim2025, saifollahi2025a}. On the other hand, methods not restricted to linear cuts have also been tested. Instead of tracing lines and polygons, it is possible to use curves that mix linear and non-linear cuts to define boundaries. For instance, \cite{saifollahi2025b} uses non-linear cuts (obtained by comparisons with artificial GCs) to select candidates in the space of magnitudes, colors, compactness index, and ellipticity; \cite{dou2025} fits ellipses to define the GC locus in the $g‐r$ vs. $r‐z$ diagram, and performs linear cuts on the morphological parameters. Another alternative is to use a statistical technique such as propensity score matching \citep{ho2007, austin2011}, so that sources are assigned as candidates based on the neighbors of each confirmed GC \citep{chies-santos2022}.
Furthermore, machine-learning models can be trained to identify GC candidates by exploring the relevant parameter space in a substantially less constrained manner than traditional approaches based on linear cuts in a small number of observables. Examples include Gaussian Mixture Models \citep{garcia-dias2020}, Random Forest classifiers \citep{Breiman2001RandomForests,biauScornett2015randomforest}, and neural networks. These techniques have recently been applied to GC identification in several studies \citep{barbisan2022, dold2022, mohammadi2022, voggel2025}. Related applications to stellar population zclassification can also be found in the context of Young Stellar Object identification \citep{Kuhn2021}. 

Regardless of the method utilized to select extragalactic GC candidates, the final selection continues to contain an important percentage of contamination from mostly background galaxies and foreground stars, which display photometric and morphological signatures similar to those of GCs.
Each of the approaches mentioned above aims to reduce contamination by making a suitable comparison between the properties of GCs from the literature and the unlabeled data (sources of unknown nature) within a parameter space defined by the available data of each study. Spatial resolution and photometry depth are crucial to decrease the number of contaminants, e.g. with HST/ACS as in \cite{jordan2015}.
For the case of the Euclid Space Telescope, by incorporating morphometric information (FWHM, ellipticity, concentration index, etc.) in the GC selection pipeline, it is expected that contamination from background sources can be reduced by 90\% \citep{voggel2025}.
Another valuable type of information that can be used to reduce contamination is kinematics. Data from the Gaia mission \citep{gaia2016b}, parallax and proper motion measurements, can help discriminate extragalactic GCs from foreground stars~\citep{hughes2021, chies-santos2022}.
In terms of SEDs and colors, it is impractical to unequivocally identify GCs from optical photometry in a few bands alone.
High contamination can be significantly reduced by including near-UV and/or near-IR information, using the $u$ and/or $K$ bands~\citep{chies-santos2011a, munoz2014, cantiello2018, saifollahi2021}. 
Ultimately, the more bands, the more extensive coverage of the electromagnetic spectrum, the more accurate information on the nature of the objects is obtained. In this sense, the definitive tool to confirm the nature of GC candidates is spectroscopy, which is, in turn, only available for a very limited number of systems, mostly nearby, with very bright GCs \citep{brodie2014}.

The task of extragalactic GC candidate selection will face an unprecedented amount of high-quality data in the next few years, with the advent of next-generation multi-band photometric surveys and telescopes such as the Vera C. Rubin Observatory's Legacy Survey of Space and Time (LSST)~\citep{lsst2019}, ESA's Euclid Space Telescope~\citep{euclidI}, 
NASA's Nancy Grace Roman Space Telescope~\citep{akesonRoman}, and the Chinese Space Station Telescope~\citep{csst2025}. 
The large volume of upcoming data requires us to adopt new approaches, in particular by looking to machine learning models.
Our work occurs in the context of preparation for Rubin/LSST; we seek to contribute to the development of tools and predictions to be applied to future LSST data.

The Vera C. Rubin Observatory is a ground-based facility with an $8.4 \ \rm m$ diameter (primary mirror) telescope, upon which the largest digital camera ever produced has been installed. The LSST camera comprises 3.2 gigapixels, with a pixel scale of $0.2 \ \text{arcsec}/ \text{pixel}$ and 9.6 square degrees of field of view.
The purpose of LSST is to iteratively map the entire southern sky to achieve very deep photometry in the 6 filters $ugrizy$\footnote{The letter $y$ is used exclusively to refer to the corresponding photometric filter present in the Rubin Observatory LSST Camera filter system. It is similar, but not identical to the $Y$ filter used in the Dark Energy Survey, which will be referred to a lot in this paper.}, with an average cadence of 3 days over 10 years of operation.
A combination of the specifications of Rubin and the LSST strategy with the literature knowledge on the GCLF (with a peak at $M_g \sim  -7.5$; \cite{jordan2007}) reveals that, with a single estimated exposure, Rubin will be able to reach the GCLF turnover magnitude (TOM) in the $g$-band of GC systems at distances up to 25 Mpc, and 0.5 mag fainter than the TOM of systems up to $\sim 20$ Mpc (about the distance to the Fornax Cluster). With coadded images over 10 years of operation, again in the $g$-band, it will reach the TOM of systems at 100 Mpc, and 1.5 mag brighter than the TOM of systems at 150 Mpc. For the entire LSST footprint, in the expected final coadded images, \cite{starClustersRoadmapLSST} estimates the detection of light from $\sim 10^{7}$ GCs in the $griz$ bands, $\sim 10^6$ in the $y$ band, and $\sim 10^5$ in the $u$-band. However, due to its limitations in terms of resolution, GCs beyond 10\,Mpc are all expected to be point-like sources in Rubin images. Therefore, it is indispensable to seek methodological improvements toward the correct identification of these remote point-like GCs that will be present in upcoming LSST data releases.

The ideal scenario would be to develop a scalable, adaptable model, 
capable of constructing clean samples of extragalactic GC candidates (resolved and unresolved) in and around a variety of galaxies, without the need for further galaxy-by-galaxy refinements, and calling external databases from other facilities (HST, Euclid, Roman, Gaia, etc.), which are available only for specific targets and restricted regions.
However, publicly available data that could be used to train and test the reliability of such a model are highly heterogeneous, which leads to non-optimal training and unreliable classifications: data from different surveys, with very different characteristics, and very few spectroscopically confirmed GCs do not allow for robust and thorough experiments in supervised setups.

Given the limitations of optical photometry for selecting GCs and the fact that Rubin lacks both near-IR filters and sufficient resolution to effectively distinguish GCs from other contaminants, the purpose of this paper is to answer the following question. With an LSST-like, $ugrizY$ photometric catalog, relying exclusively on color information\footnote{As a starting point, we aim to understand the specific role of Rubin/LSST colors for the identification of GCs. Thus, for the time being, we abstract away from SEDs and morphometric measurements, considering them as additional information to eventually complement the color space.}, how well point-like extragalactic GCs can be distinguished from background galaxies and foreground stars (the main contaminants)? 
An equivalent way of framing our goal is: do traditional color-color diagrams capture all of the clustering capability available in such a multi-band photometric catalog? If not, i.e., assuming the color space in the catalog does contain sufficient information to further distinguish point-like GCs from contaminants (beyond what is achieved with 2D color-color diagrams), then there must exist a transformation capable of making this information evident, enhancing the contrast between the different classes of objects. In that context, we investigate whether this is the case or not.

Toward our goal, we assemble an LSST-like, photometric catalog containing labeled confirmed GCs from the literature, 
as well as background galaxies and foreground stars, as described in Section~\ref{sec:data}.
Then, as detailed in Section~\ref{sec:methods}, we transform the colors available in this dataset in two different ways to address the questions posed above, (i) using principal component analysis \citep[PCA;][]{jolliffe2016} and (ii) training non-linear auto-encoders \citep[AEs;][]{bank2020autoencoder}. For our purposes, these techniques serve as dimensionality reduction tools \citep[see e.g.,][]{Xu2023,qrpca,rpca2014}. 
We then use machine learning classification models to test whether these transformations on the colors can improve the identification of the GCs in comparison with the use of the colors themselves as input to the models.
The results of our tests are presented and discussed in Section~\ref{sec:results}.
We summarize our results and suggestions in Section~\ref{sec:conclusions}.

\section{Data}\label{sec:data}
Since LSST will provide data in the six bands $ugrizy$, we choose publicly available multi-band photometry data of the Fornax Cluster provided by the Dark Energy Survey (DES) in the $grizY$ bands~\citep{abbottDES-DR2-2021}, alongside ESO/VST's (European Southern Observatory's VLT Survey Telescope) Fornax Deep Survey (FDS), in the $ugri$ bands~\citep{cantielloFDS2021}, to compose an LSST-like photometric catalog.

\subsection{Fornax Deep Survey data and confirmed globular clusters}\label{subsec:fds}
The Fornax Deep Survey is a deep imaging survey performed with the ground-based ESO VST, 
a $2.6 \ \rm  m$ diameter telescope at Cerro Paranal, Chile.
It used the OmegaCAM camera to obtain images of 26 square degrees of the Fornax Cluster in the four bands $ugri$, 
with a pixel scale of $0.21 \ \text{arcsec} \ \text{pixel}^{-1}$, and a field of view of 1.0 square degrees~\citep{peletier2020}. 
The FDS photometry described in \cite{cantielloFDS2021} uses a multi-band coadded image created by stacking the best quality (sharpest FWHM) coadded single-band images in $gri$. \texttt{SExtractor} receives this stack image as input to derive properties such as the mean FWHM, \texttt{CLASS\_STAR}, and flux radius. Moreover, \texttt{DAOphot} is used in this image to model the PSF and identify sources. Magnitudes are then estimated by integration over the PSF in each single-band coadded image. We access the resulting catalogs through the VizieR catalog access tool \citep{vizier2000}.
All the magnitudes in question are in the AB system.
Important quantities about FDS observations and photometry are presented in Table~\ref{tab:fds}.

\begin{table}[h]
    \centering
    \begin{tabular}{lcccc}
     & $u$ & $g$ & $r$ & $i$ \\
    \hline
    Magnitude limit (mag) & 24.1 & 25.4 & 24.9 & 24.0 \\
    PSF FWHM (arcsecond) & 1.26 & 1.12 & 0.92 & 0.94 \\
    \hline
    \end{tabular}
    \caption{FDS data quality information from \cite{cantielloFDS2021}. 
             The first row reports limiting magnitudes derived from $5\sigma$ magnitude integration over the PSF. The second row presents the median FWHM.}
    \label{tab:fds}
\end{table}

Furthermore, \cite{cantielloFDS2021} matched the FDS $ugri$ photometric catalog with the spectroscopic samples of GCs in Fornax produced by \cite{schuberth2010} and the Fornax Cluster VLT Spectroscopic Survey \citep{pota2018}. In this way, 1342 spectroscopically confirmed GCs, the majority associated with NGC 1399, are labeled in the FDS catalog. Additionally, another 1921 sources are labeled as photometrically confirmed GCs due to their marginally resolved appearances in the ACS Fornax Cluster Survey (ACSFCS) images \citep{jordan2007}, located in other galaxies across the Fornax Cluster. As described in \cite{cantielloFDS2021}, these 1921 sources are selected from a cut on the probability of being a GC $p_{\rm GC} > 0.75$; see \cite{jordan2009} for a detailed definition of this probability measure. Still, of these photometrically confirmed GCs, 214 have $p_{\rm GC}$ below 95\%, 110 below 90\%, and only 23 below 80\%. Finally, \cite{chaturvedi2022} increased the number of spectroscopically confirmed GCs in the Fornax Cluster central region. Out of their catalog of confirmed GCs, 296 were previously unidentified in the catalog of all FDS sources from \cite{cantielloFDS2021}, hence we use FDS IDs to label them accordingly. In this work, we take into consideration all of the 1342 + 268 spectroscopically confirmed GCs, together with the 1921 photometrically confirmed ones, hereafter referred to as simply ``confirmed GCs''.

\begin{figure}
    \centering
    \includegraphics[width=1.0\linewidth]{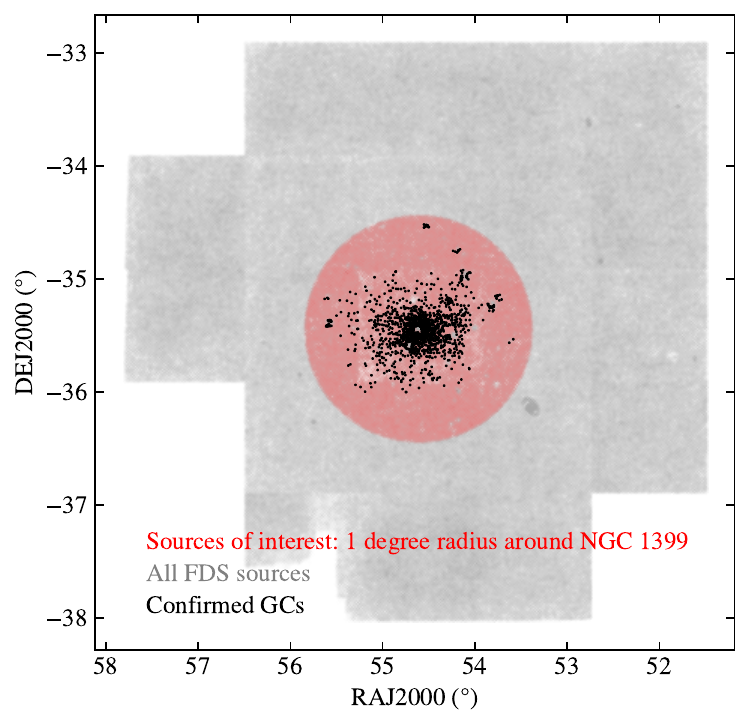}
    \caption{Entire coverage of FDS in gray, the red circle indicates the 1-degree radius region around NGC 1399; our sources of interest. The black points represent the positions of confirmed globular clusters for which we have FDS $ugri$ and DES $grizY$ photometry available.}
    \label{fig:spatialDistribution}
\end{figure}

As the scope of this paper is to compare the effects of different methods of transforming colors to identify GCs, rather than to select new GC candidates, we restrict our analysis to the sources found within the circular region with radius of 1 degree around NGC 1399, within which the photometric calibration across bands (and therefore colors) are the most homogeneous. This region contains all of the spectroscopically confirmed GCs mentioned above and a couple hundreds of the photometrically confirmed ones. See Figure~\ref{fig:spatialDistribution} for a visualization of the spatial distribution of the sources. Given the instrumental details of FDS and the distance to the Fornax Cluster ($\sim 19.3$ Mpc; \cite{anand2024}), GCs are point-like sources \citep{cantielloFDS2021}.

\subsection{Dark Energy Survey data and the selection of stars and galaxies}\label{subsec:des}
The Dark Energy Survey (DES) is a ground-based, wide-area visible and near-infrared imaging survey that covers
approximately 5000 square degrees of the southern sky in the $grizY$ bands.
DES imaging is performed with the Dark Energy Camera (DECam), which is mounted on the $4 \ \rm m$ Blanco Telescope at 
Cerro Tololo Inter-American Observatory in Chile \citep{abbottDES-DR2-2021}, with a pixel scale of $0.264~\text{arcsec} \  \text{pixel}^{-1}$ and a field of view of 3 square degrees.
We use the DES DR2 photometric catalog obtained via \texttt{SExtractor}~\citep{bertin1996} in double-image mode, where the detection image is the linear combination stack of coadded images in the three bands $r+i+z$. DES DR2 contains magnitude estimations using several aperture models. The most important for this work are \texttt{MAG\_AUTO} (elliptical model based on the Kron radius) and \texttt{MAG\_APER} (circular apertures). We disregard DES weighted-average PSF photometry upon realizing that, for the 1 degree radius circular region around NGC 1399, it contains large fractions of missing values across all $grizY$ bands; respectively, $52,43,51,63,80\%$. Such high fractions are a consequence of the fact that weighted-average PSF magnitudes are measured only for sufficiently bright sources that are detectable in single-epoch images \citep{abbottDES-DR2-2021}.
All DES magnitude values are in the AB system.

We access DES data through the Astro Data Lab science platform \citep{nikutta2020} and, as explained in Subsection~\ref{subsec:fds}, we select sources that lie within the circular area of 1 degree radius around NGC 1399, which are shown in Figure~\ref{fig:spatialDistribution}; a total of 395813 sources.
Some relevant quantities on the DES DR2 data are presented in Table~\ref{tab:des}. As they are in the FDS images, GCs in the Fornax Cluster are point-like sources in the DES DR2 images.
\begin{table}[h]
    \centering
    \begin{tabular}{lccccc}
        & $g$ & $r$ & $i$ & $z$ & $Y$ \\
    \hline
    Magnitude limit (mag) & 24.7 & 24.4 & 23.8 & 23.1 & 21.7 \\
    PSF FWHM (arcseconds) & 1.11 & 0.95 & 0.88 & 0.83 & 0.90 \\
    \hline
    \end{tabular}
    \caption{DES DR2 data quality information from \cite{abbottDES-DR2-2021}. 
             The first row displays the median limiting magnitudes of the coadded catalog for a $1.95''$ 
             diameter aperture (\texttt{MAG\_APER\_4}), at $S/N=10$.}
    \label{tab:des}
\end{table}

Beyond labeling spectroscopically confirmed GCs and joining DES and FDS photometric catalogs, we also have to label samples of background galaxies and foreground stars. The selection criteria must ensure minimum contamination from unidentified GCs, which could otherwise compromise the interpretation of the performance of the classification models later on.
Towards this end, we leverage the galaxy-star separation criteria suggested by \cite{abbottDES-DR2-2021} to select clean samples of galaxies and stars from DES DR2 data. These criteria combine a morphology-based classification variable named \texttt{EXTENDED\_COADD}, which in turn is based on \texttt{SExtractor} \texttt{SPREAD\_MODEL}, with the magnitude values \texttt{mag\_auto\_i}. We use altered versions of the selection criteria presented in \cite{abbottDES-DR2-2021}, considering now fixed values of \texttt{EXTENDED\_COADD} in each selection, and modifying the \texttt{mag\_auto\_i} cut from [19.0, 22.5] to [18.0, 20.0] for the stellar selection:
\begin{eqnarray}
    \text{Galaxy selection: }&  \texttt{EXTENDED\_COADD} = 3 \;\;\& \;\; 19.0 \le \texttt{mag\_auto\_i} \le 22.5 \label{eq:galaxy} \\
    \text{Stellar selection: }& \texttt{EXTENDED\_COADD} = 0 \;\;\& \;\; 18.0 \le \texttt{mag\_auto\_i} \le 20.0 \label{eq:star} 
\end{eqnarray}
The values of 3 and 0 for \texttt{EXTENDED\_COADD} correspond to high-confidence galaxies and stars, respectively, whereas the original criteria consider the ranges $\texttt{EXTENDED\_COADD} \ge 2$ and $0 \ge\texttt{EXTENDED\_COADD} \ge 1$ to include the ``likely'' galaxies and stars in the selections. We refer to Subsection 4.5 of \cite{abbott2018} for a detailed description of the connection between \texttt{EXTENDED\_COADD} and \texttt{SPREAD\_MODEL}, and of the reason why such variables may render more reliable morphological classifications compared to using \texttt{CLASS\_STAR}.

We now discuss the motivation behind the stellar selection criteria modification, as well as the possible contamination from unlabeled GCs in our galaxy and star samples.
As the GCs in question are all point-like sources, \texttt{EXTENDED\_COADD} must provide enough information to select a very pure sample of extended background galaxies; no contamination from point-like sources is expected in this case.
Regarding our stellar sample, the original selection criterion from \cite{abbottDES-DR2-2021} considers $19.0 \leq \texttt{mag\_auto\_i} \le 22.5$, which arguably leads to an appreciable portion of point-like GCs in Fornax being selected as stars. Namely, at the distance of Fornax, $m-M \sim 31.5$, and assuming an absolute GCLF TOM in the $i$-band $\text{TOM}_i \sim -8.5$ mag, we obtain an apparent $\text{aTOM}_i \sim 23$ mag. Now, with a GCLF spread $\sigma_{\rm GCLF} \sim 1.5$ mag for the entire Fornax cluster (this accounts for the cluster depth), we get that the cut $\texttt{mag\_auto\_i} \le 22.5$ is at $\sim 0.33 \sigma_{\rm GCLF}$ from $\text{aTOM}_i$. This implies that the stellar sample would still contain roughly 35 -- 40\% of the entire GC population in Fornax. This fraction is reduced to $<3\%$ if the cut is at 20 mag.
This is why we decided to modify this cut, as a way to ensure that unidentified GCs represent a very small, negligible fraction of the total sample of stellar objects. We also shift the bright-end of the cut from 19.0 to 18.0 so that the selection is not too restrictive.
The actual selection of galaxies and stars is performed with an extra \textit{ad hoc} constrain: exclude all sources currently labeled as confirmed GCs. The final number of selected galaxies and stars is discussed in Subsection \ref{sub:preproc}.

\cite{abbottDES-DR2-2021} also states that, following criteria~\ref{eq:galaxy} and the original version of \ref{eq:star}, galaxies and stars can be selected with efficiency rates greater than $99\%$ and $94\%$, and contamination rates lower than 2\% and 3\%, respectively. We stress that such values are derived in relation to the entirety of HSC SSP DR1~\citep{aihara2018}, and thus do not imply that the use of the \texttt{SPREAD\_MODEL} is sufficient to break nearly all degeneracies in the identification of compact objects such as GCs, as discussed above.

After cross-matching FDS and DES catalogs, we apply criteria \ref{eq:galaxy} and \ref{eq:star} to label extended background galaxies and point-like foreground stars in the resulting matched catalog to be described in the two following Subsections.

\subsection{LSST-like dataset: combining DES and FDS data}\label{subsec:lsstlike}
As described in Section~\ref{sec:intro}, LSST will measure the light of an enormous number of extragalactic GCs; it will detect GCs out to $\sim 200$ Mpc \citep{starClustersRoadmapLSST}. 
The expected LSST limiting magnitudes for single images and coadded ones (after 10 years of observations) are shown in Table~\ref{tab:lsst}.
\begin{table}[h]
    \centering
    \begin{tabular}{lcccccc}
        & $u$ & $g$ & $r$ & $i$ & $z$ & $y$ \\
    \hline
    Single image lim. magnitude (mag) & 23.8 & 24.5 & 24.0 & 23.4 & 22.7 & 22.9 \\
    Coadded image lim. magnitude (mag) & 25.6 & 26.9 & 26.9 & 26.4 & 25.6 & 24.8 \\
    \hline
    \end{tabular}
    \caption{The expected $5\sigma$ depths for LSST single exposure and coadded images (after the 10 years survey), estimated from operation simulations~\citep{bianco2022}.}
    \label{tab:lsst}
\end{table}

Comparing Tables~\ref{tab:fds}, \ref{tab:des}, and \ref{tab:lsst}, we see that FDS achieves magnitude limits deeper than DES, while both are deeper than the expected limiting magnitudes of single exposures of LSST, except in the $y$-band. However, the expected photometric depths of coadded images after 10 years of the LSST survey are well beyond those of FDS and DES. Moreover, the values of the pixel scales and median FWHMs of sources in DES and FDS are very similar, suggesting that the combination of their photometric data should not create compromising biases in terms of data quality. The details of this combination are now discussed.

First, we perform a cross-match with the sky coordinates in the DES and FDS catalogs. An angular separation tolerance of $1.0''$ was used based on the median FWHM values of the surveys. Of the 1342 spectroscopically confirmed GCs present in the catalog of \cite{cantielloFDS2021}, 1129 are also available in DES DR2 catalog. Of the 292 spectroscopically confirmed GCs from \cite{chaturvedi2022}, after filtering out those whose spectra have $S/N < 3$, 268 are also present in the DES DR2 catalog. Finally, 245 photometrically confirmed GCs are in the matched catalog, giving us a total of 1642 confirmed GCs at this stage. The missing GCs in question are all located within the very bright central region of their host galaxies (including NGC 1399), which are disregarded by the source extraction and deblending routines of the DES pipeline \citep{abbott2018}.
It is important to emphasize that the strategy used to extract and measure the sources in the FDS catalog was designed toward compact systems science, with the goal to study globular clusters and ultra-compact dwarf galaxies. Meanwhile, DES aims to produce consistent measurements over a large number of observations made throughout $\sim 5000$ square degrees of the southern sky. This is consistent with the number density of sources in the Fornax Cluster being greater in FDS than in DES. All mentions to ``catalog data'' hereafter refer to the matched catalog.

Although only FDS has data in the $u$-band, and the $z$ and $Y$ bands are available only in DES, the two surveys share the $gri$ bands. Beyond that, DES and FDS provide magnitudes using various aperture models. To decide which magnitude values to use for our analysis, it is necessary to further compare the photometry from FDS and DES as follows. All magnitude values in both FDS and DES were corrected by reddening in accordance with \cite{schlegel1998} and \cite{schlafly2011}.

\begin{figure}
    \centering
    \includegraphics[width=1.0\linewidth]{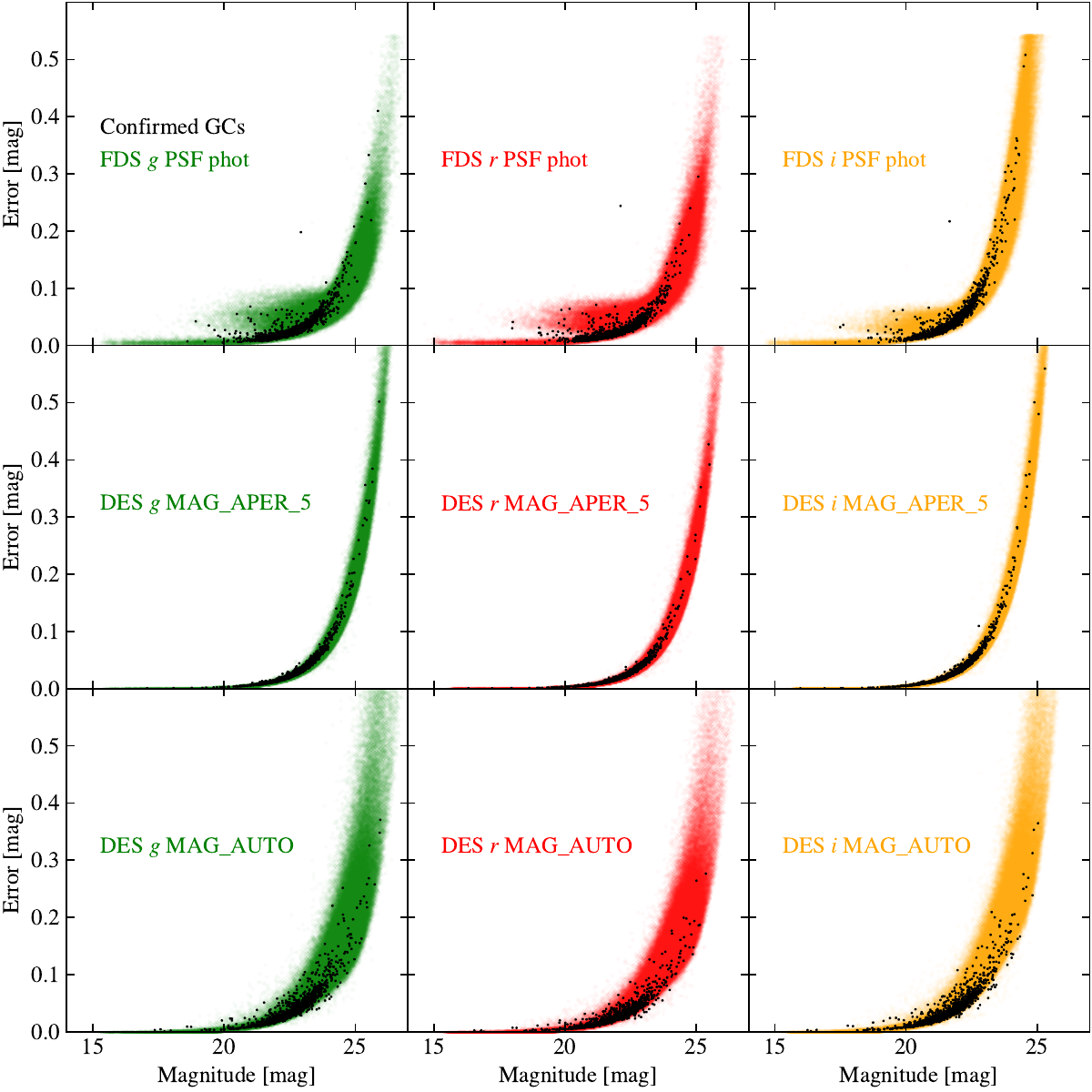}
    \caption{Magnitude errors plotted versus magnitudes for the bands in common between DES and FDS, $gri$. Black dots represent confirmed GCs. The first row of plots refers to FDS PSF photometry data, the second to DES circular aperture photometry, and the third to DES automatic aperture (based on the Kron radius) photometry. For visualization purposes, the magnitude error axes were truncated at a value of $0.6$ mag. Errors in FDS data do not exceed $0.6$ mag: all FDS data points are visible in these plots.}
    \label{fig:magVSmagError}
\end{figure}

\begin{figure}
    \centering
    \includegraphics[width=1.0\linewidth]{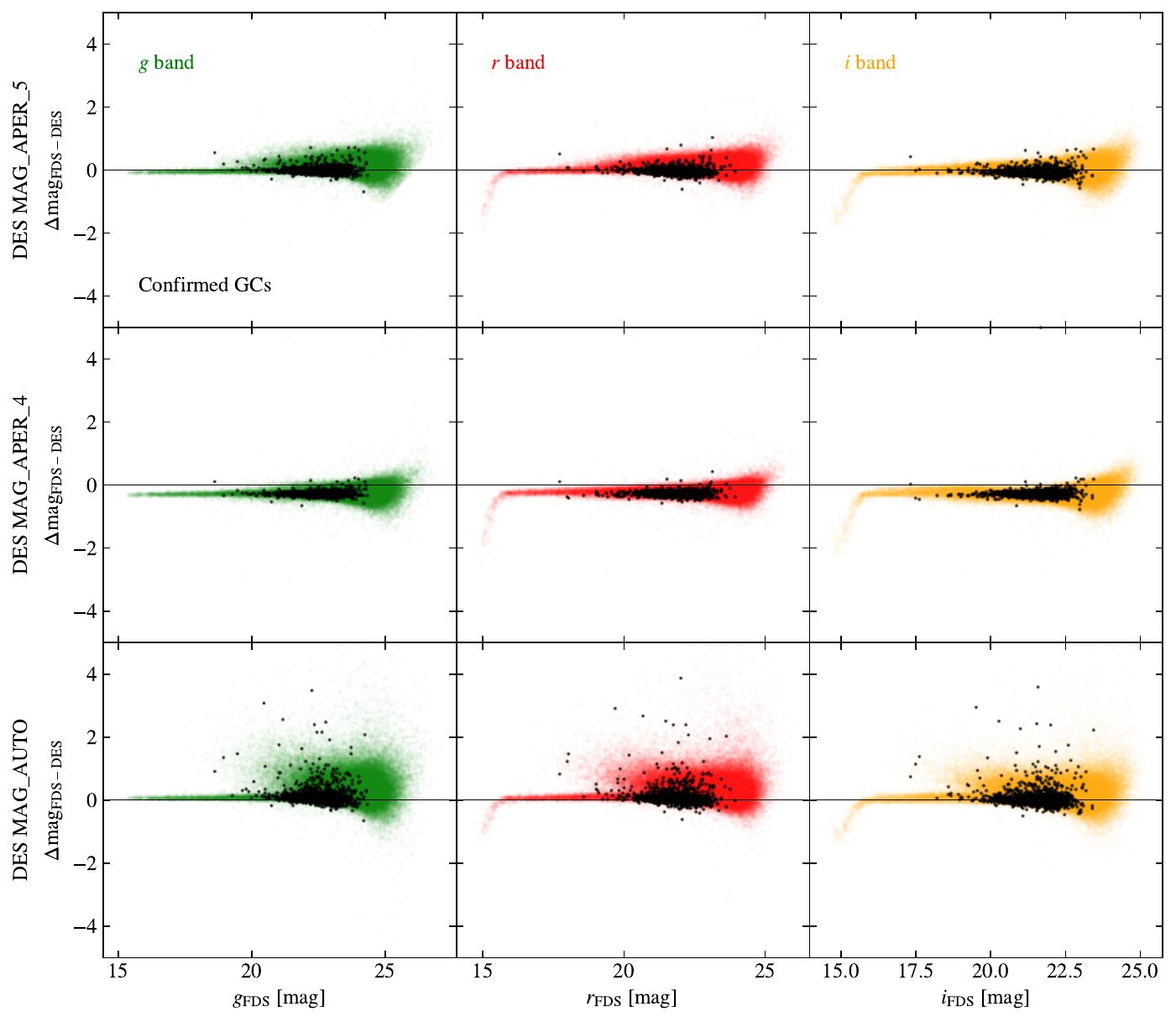}
    \caption{$\Delta\rm{mag_{FDS-DES}}$ versus $g,r,i_{\rm FDS}$: the difference in magnitude for the same source, in the same band, but in different surveys against the FDS magnitude in the same band. The first row displays the plots where DES \texttt{MAG\_APER\_5} ($2.92''$) data was used, the second DES \texttt{MAG\_APER\_4} ($1.92''$), and the third DES \texttt{MAG\_AUTO}. The horizontal black line is $y=0$.}
    \label{fig:magDiff}
\end{figure}

We first consider only the $gri$ bands, available in both surveys. In Figure~\ref{fig:magVSmagError}, the thinnest exponential scatters indicate that DES circular aperture photometry has the lowest errors among the other models for these three bands. The spectroscopically confirmed GCs also have the lowest magnitude errors in the circular aperture model. \texttt{MAG\_APER\_5} corresponds to $11.11$ pixels or $2.92$ arcseconds of diameter, but we also studied the error curve for smaller circular apertures of $1.95$ and $1.46$ arcseconds (\texttt{MAG\_APER\_4} and \texttt{MAG\_APER\_3}, respectively). They are qualitatively indistinguishable from those presented in the middle row of Figure~\ref{fig:magVSmagError} and thus are not included in the figure. Furthermore, \texttt{SExtractor} magnitude error estimation is found to be more reliable compared with that of \texttt{DAOphot} \citep{chies-santos2011a}.
Therefore, from the perspective of the magnitude error curves, the circular aperture model is favored over the automatic one. Still, there are many options for aperture size.

To choose the most suitable aperture size, we examine Figure~\ref{fig:magDiff}, which shows the differences in magnitude values for the same sources, in the same band, but in different surveys and different aperture models and sizes. We find that DES \texttt{MAG\_APER\_5} and \texttt{MAG\_AUTO} are the models whose magnitudes deviate the least from the PSF photometry of FDS (taken here as reference), especially when considering sources brighter than 20 mag in the three bands. However, for fainter sources, DES \texttt{MAG\_AUTO} values differ a lot more from the FDS magnitudes compared to the circular aperture models, which are more consistent. For DES \texttt{MAG\_APER\_4} magnitudes, a vertical shift is observed: with this circular aperture, DES magnitudes are systematically fainter than FDS'. For DES \texttt{MAG\_APER\_5} this shift is not as pronounced; that is, a larger aperture accounts for light that was not captured in with \texttt{MAG\_APER\_4}, but was picked by the PSF photometry of FDS. Statistics on these magnitude differences are presented in Table~\ref{tab:magDiffStats}. 
\begin{table}[h]
    \centering
    \begin{tabular}{ccccc}
    Filter & Mean $\Delta\rm{mag_{FDS-DES}}$ (mag) & Median $\Delta\rm{mag_{FDS-DES}}$ (mag) & RMS $\Delta\rm{mag_{FDS-DES}}$ (mag) & $\sigma_{\rm FDS}$ \\
    \hline
    $g$ & (0.023,  $-$0.318, 0.184)  & (0.030,  $-$0.308, 0.141)  & (0.32, 0.40, 0.54) & 1.26 \\
    $r$ & (0.023,  $-$0.272, 0.182)  & (0.037,  $-$0.264, 0.147)  & (0.28, 0.38, 0.48) & 1.40 \\
    $i$ & ($-$0.042, $-$0.370, 0.052)  & ($-$0.043, $-$0.340, 0.074)  & (0.40, 0.48, 0.51) & 1.44 \\
    \hline
    \end{tabular}
    \caption{Statistics on the differences in magnitude values for the bands in common between FDS and DES. The values are displayed in triplets representing the different aperture models for DES photometry, while FDS PSF magnitudes are the same for each comparison: (\texttt{MAG\_APER\_5}, \texttt{MAG\_APER\_4}, \texttt{MAG\_AUTO}). $\sigma_{\rm FDS}$ is the standard deviation in FDS magnitudes for reference.}
    \label{tab:magDiffStats}
\end{table}
Among the DES aperture models studied compared to FDS PSF photometry, the lowest means, medians, and RMSs are found to be those associated with DES \texttt{MAG\_APER\_5}. We performed the same examination also considering smaller and larger circular apertures (e.g. \texttt{MAG\_APER\_3} $\sim1.46''$ and \texttt{MAG\_APER\_6} $\sim3.90''$) and the results are as expected: smaller apertures disregard even more light compared with the case of \texttt{MAG\_APER\_4}, yielding even greater magnitude difference means, medians, and RMSs. Meanwhile, apertures larger than that of \texttt{MAG\_APER\_5} also cause these statistical measures to increase in value, likely because too large apertures increase noise and/or capture part of the light of neighboring sources. Finally, the distributions of spectroscopically confirmed GCs for the DES \texttt{MAG\_APER\_\{5,4\}} cases in Figure~\ref{fig:magDiff} are more concentrated around the $y=0$ line when compared with the \texttt{MAG\_AUTO} case, once again favoring circular aperture over the automatic one. 

Based on the investigation above, we decided to compose our LSST-like $ugrizY$ photometric catalog with the $u$ band from FDS PSF photometry alongside DES $grizY$ \texttt{MAG\_APER\_5} circular aperture photometry (11.11 pixels or $2.92''$ of diameter). No aperture correction was applied.

We are interested not only in confirmed GCs, but also in labeling background galaxies and foreground stars using criteria \ref{eq:galaxy} and \ref{eq:star}. Thus, we include the same plots as in Figure~\ref{fig:magDiff} in Appendix~\ref{app:data}, but highlighting these two categories of contaminants in Figures~\ref{fig:magDiffGalaxiesInBlack} and~\ref{fig:magDiffStarsInBlack}. 

\subsection{Data pre-processing: filtering and labeling sources}
\label{sub:preproc}

Another important aspect to consider is the presence of missing values. 
In our catalog, about 9.0\% of the data points miss the $Y$-band magnitude, and less than 1\% miss magnitude values in other bands. The FDS $u$ band does not have any missing values. We removed any sources that lack a magnitude value in at least one band, leaving us with 190281 sources, of which 1595 are confirmed GCs.

\begin{figure}
    \centering
    \includegraphics[width=1.0\linewidth]{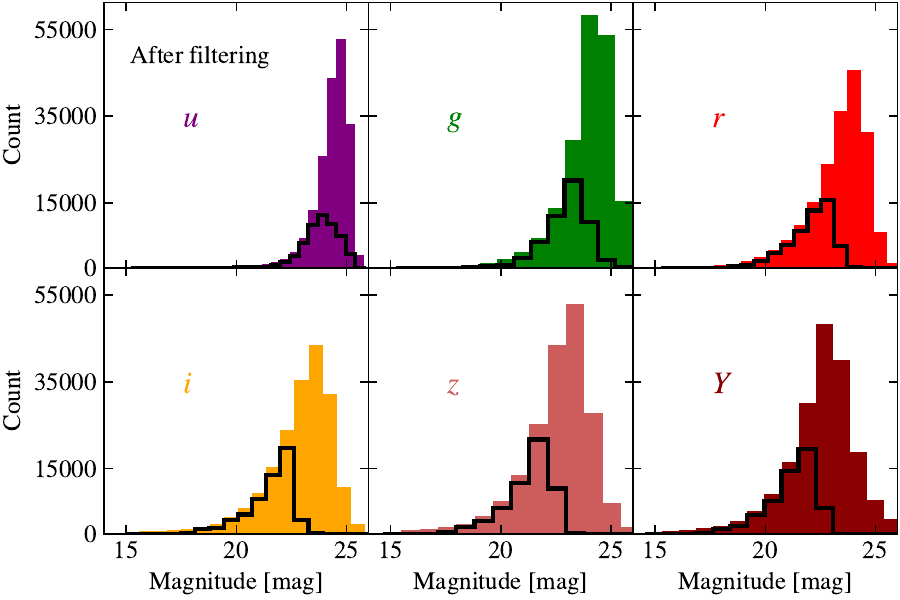}
    \caption{Distribution of magnitude values for each band. The colored bars represent the dataset before the pre-processing described in Subsection~\ref{sub:preproc}. The black edges indicate the subset that represents the dataset after all the filtering; it contains only the labeled sources.}
    \label{fig:magHist}
\end{figure}

We filter out all sources that have at least one of the $ugrizY$ bands with a magnitude error greater than 0.5 mag. This procedure leaves us with 105318 sources in total.
Lastly, we add labels to all the remaining sources that fall within the DES selection criteria \ref{eq:galaxy} and \ref{eq:star}. These sources, together with the confirmed GCs, comprise our final, LSST-like, filtered and labeled dataset with three classes: 1440 confirmed GCs, 49579 background galaxies, and 3726 foreground stars; 54745 sources in total. The fact that the different classes have very different numbers of associated data points (class imbalance) is taken into account when training the classification models. Figure~\ref{fig:magHist} allows us to visualize the scale of the filtering we applied to the data. 

We recall the discussion on our efforts to minimize the contamination from unidentified GCs in our samples galaxies and stars, presented in \ref{subsec:des}. Background galaxies indeed dominate the total number of sources in our curated catalog, although no point-like GCs nor stars are expected to be present in this sample due to the role of \texttt{EXTENDED\_COADD} in criterion \ref{eq:galaxy}. The magnitude cut in criterion \ref{eq:star}, together with our knowledge of the GCLF of the Fornax Cluster, led us to expect unknown GCs to represent less than 3\% of our stellar sample. In addition, we ran the TRIdimensional modeL of thE GALaxy population synthesis code (TRILEGAL; introduced and calibrated by \cite{girardi2005}) using its default parameter values to obtain an estimate for the number of galactic stars within the 1-degree circular region around NGC 1399 from synthetic apparent magnitudes in the DES filters. After imposing the DES magnitude limits to filter our TRILEGAL output catalog and applying the same magnitude cut as in criterion \ref{eq:star}, we arrive at an estimated total of 4756 galactic stars within the sky region in question, which differs from the number of foreground stars in our sample by about a thousand counts. We interpret this difference to be barely within the expected uncertainty in star counts predicted by TRILEGAL, of about 20\% \citep{dalTio2022}, placing our foreground star sample in borderline agreement with the model regarding star counts.

\section{Methodology}
\label{sec:methods}

Given the filtered and labeled dataset, we formulate the task as a classification problem aimed at assessing whether color transformations via dimensionality reduction enhance the separation of the confirmed GCs from the select background galaxies and foreground stars.
We evaluate this using two classifiers: a random forest \citep[RFC;][]{Breiman2001RandomForests} and a multi-layer perceptron  \citep[MLPC;][]{murtagh1991}.

We consider both RFCs and MLPCs to probe the extent to which the information content of the LSST-like catalog requires non-linear decision boundaries. RFCs provide a strong, well-calibrated baseline for tabular data, capturing feature interactions with minimal tuning and offering robustness to noisy photometry and heterogeneous source populations. MLPCs, by contrast, impose fewer structural assumptions and can represent smoother, higher-capacity non-linear mappings in color space. Comparing their performances, therefore, serves as a diagnostic: a consistent gain from MLPCs would suggest that the class separation benefits from higher-capacity, continuous non-linear representations, whereas comparable performance would indicate that the discriminative signal is largely captured by lower-capacity tree ensembles (or is limited by measurement noise, selection effects, and label uncertainty) rather than by model expressiveness.

These models are trained and tested using three types of input: (i) colors derived from $ugrizY$ photometry, (ii) the principal components (PCs) of these colors, and (iii) the latent space coordinates (LSCs) of auto-encoders (AEs; \cite{bank2020autoencoder}) that also take the colors as input. For our purposes, AEs can be understood as neural networks designed to compress data efficiently, a dimensionality reduction tool, but not limited to linear transformations \citep{fournierAloise2021}, which is the case of PCA. For a more detailed description of AEs, see Subsection~\ref{subsec:inputs}. 

The core idea behind using PCA and AEs to transform the colors is that both techniques can potentially compact information into lower-dimensional representations, revealing and emphasizing clustering patterns in the data. This idea was also explored by, e.g., \cite{dabrusco2016} and \cite{chies-santos2022}. \cite{dabrusco2016} selected GC candidates around NGC 1399 using PCA on a 3D color space of FDS catalogs, and reported its GC system to be extended so that it connects with those of neighboring galaxies. \cite{chies-santos2022} selected GC candidates in the M81/M82/NGC3077 triplet using the PC1-PC2 diagram derived from the 12-dimensional SED space of J-PLUS photometric catalogs. Although no quantitative comparison with other methods was presented in these studies, they obtained solid lists of GC candidates, with their spatial distributions being interpretable in light of cluster-galaxy and galaxy-galaxy interactions.
We aim to provide such a quantitative comparison.

Our choice to consider exclusively the color space instead of the combined space of SEDs and colors is justified as follows. During our preliminary tests, we noted that the use of colors as the only input (as opposed to using colors and SEDs) yields a slightly more compact distribution of GCs in principal component diagrams. In the context of our specific investigation, we interpreted this as an indication that the SEDs in question (each composed of only 6 data points) do not carry information capable of improving the clustering of GCs already observed in the color space, or in the space of PCs of colors. Furthermore, associated with the above is the fact that our samples of labeled objects (GCs, background galaxies, and foreground stars) are biased towards brighter sources. Therefore, choosing inputs to be distance-independent quantities such as colors also serves as a way to mitigate this selection bias.

\begin{figure}
    \centering
    \includegraphics[width=1.0\linewidth]{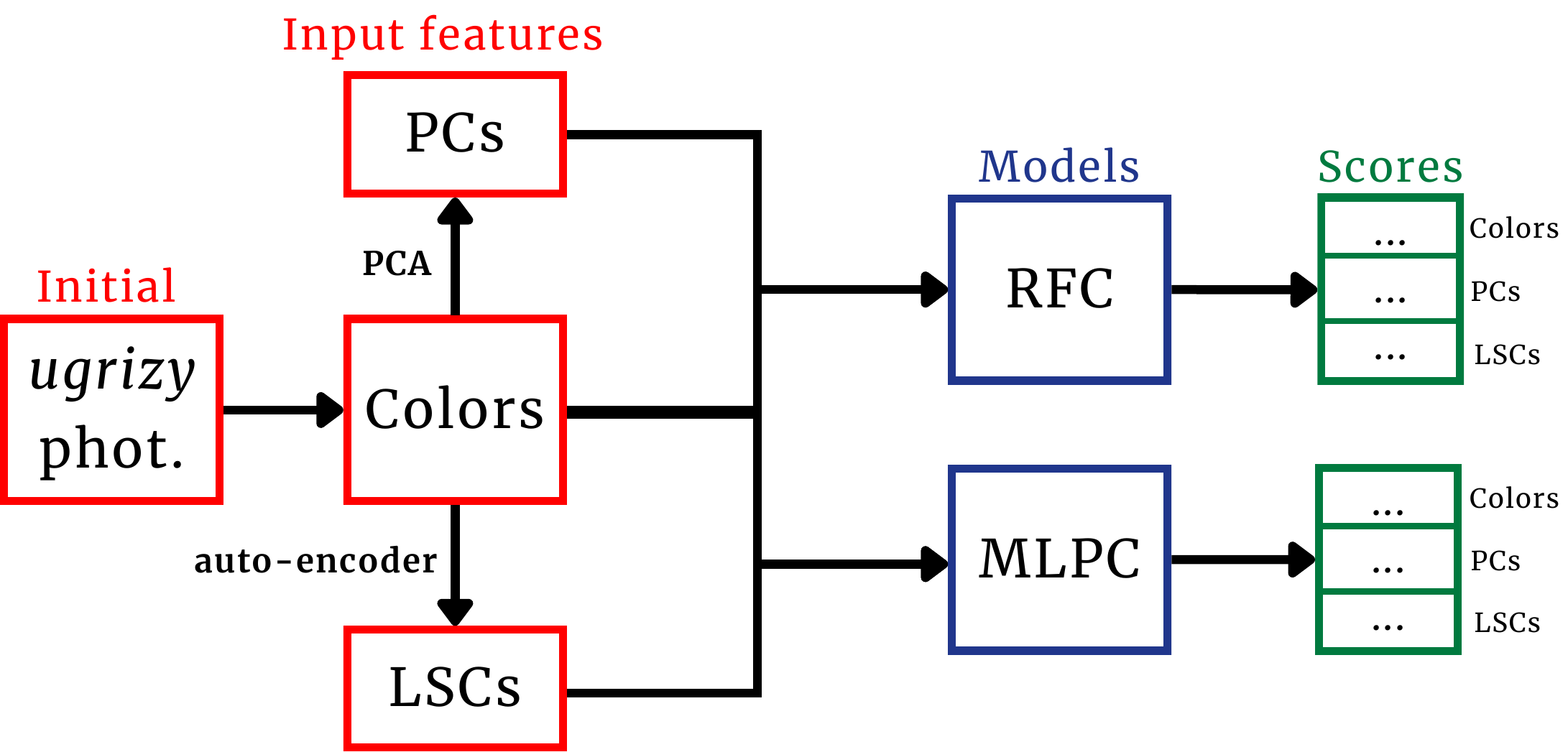}
    \caption{Diagram to illustrate the flux of data in our analysis procedure. PCs $\sim$ principal components, LSCs $\sim$ latent space coordinates, RFC $\sim$ random forest classifier, MLPC $\sim$ multi-layer perceptron classifier.}
    \label{fig:schematic}
\end{figure}

To evaluate the performance of the models, we use the output metrics precision (Eq.~\ref{eq:precision}), recall (Eq.~\ref{eq:recall}), and F1-score (Eq.~\ref{eq:f1}). In their expressions, TP stands for true positives, FP refers to false positives, and FN to false negatives. 
Figure~\ref{fig:schematic} synthesizes our analysis procedure.

\begin{eqnarray}
    \rm P =& \rm TP / (TP + FP) \label{eq:precision} \\
    \rm R =& \rm TP / (TP + FN) \label{eq:recall} \\
    \rm F1 =& 2 \rm PR / (\text{P} + \text{R}) \label{eq:f1}
\end{eqnarray}

\subsection{Preparing model inputs}\label{subsec:inputs}
We aim to investigate whether more concise inputs could affect the identification of GCs. For this, we prepare input sets to the models using different numbers of colors, PCs, and LSCs\footnote{This is equivalent to stating that we are interested in assessing the effects of dimensionality reduction on our classification problem.}.

\subsubsection{Colors}\label{subsubsec:color}
The available colors are obtained by subtracting two magnitude values in different bands in all possible ways. 
We are working with the $ugrizY$ filter-set that comprises 15 colors: 
$u-g$, $u-r$, $u-i$, $u-z$, $u-Y$, $g-r$, $g-i$, $g-z$, $g-Y$, $r-i$, $r-z$, $r-Y$, $i-z$, $i-Y$, $z-Y$.
We define 3 input sets to both RFC and MLPC, composed of colors only.
The first one comprises all 15 available colors as input. Among other outputs, RFCs provide a feature ``importance'' measurement, which uses internal metrics such as Gini importance or Mean Decrease in Accuracy to quantify how much each input feature contributes to the prediction accuracy of the model. Using the feature importance values obtained when running the RFC with 15 colors, we choose the 4 ``most important'' colors, which were $u-g$, $u-z$, $u-i$, and $g-r$, to constitute our second color input set, a 4D input to both RFC and MLPC. 
Finally, using the same RFC feature importance ranking, the 2 most important colors, $u-g$ and $u-z$, are the third color input set.

\subsubsection{Principal components}\label{subsubsec:pc}
The PCs of the colors are obtained via the Principal Component Analysis (PCA) class implemented in the Scikit Learn Python package, which, in turn, uses singular value decomposition to compute the PCs \citep{scikit-learn}.
To suitably compare the model runs using colors and PCs as input to RFC and MLPC, we again consider three input sets: all 15 PCs, the first 4 PCs, and the first 2 PCs.

\subsubsection{Latent space coordinates}\label{subsubsec:lsc}
We now discuss AEs and their latent spaces \citep{bank2020autoencoder}. 
The purpose of an AE, which is a neural network, is to compress, i.e., to encode the input data into a lower-dimensional representation (the latent space, represented by the innermost layer of the network) such that it contains enough information to reconstruct the original data up to a user-defined acceptable loss. In this sense, an AE is a dimensionality reduction tool \citep{fournierAloise2021}.
The part of the network responsible for the compression is called the encoder, 
while the one used to reconstruct the original data is named the decoder. 
Naturally, the target data used to compute the loss function and train the network must be identical to the input data: one desires to minimize the difference between the input data and its reconstructed version from the lower-dimensional latent space to obtain the best compression possible. 
The dimension of the latent space, that is, the dimension of the compressed version of the data, is to be defined by the user.
Unlike the case of PCA, which outputs 15 linear combinations if 15 colors are given, for an AE, a latent space with as many dimensions as the input data is meaningless because no compression is achieved. 

We decided to use two AEs, one with a latent space of 4 dimensions and the other with 2, thus transforming the set of 15 colors into 4D and 2D representations. Again, this choice allows for more direct comparisons with the other 4D and 2D inputs composed of colors and PCs.
Therefore, we make two runs with the coordinates of the colors in the latent spaces (LSCs) as input to each classification model.

\begin{figure}
    \centering
    \includegraphics[width=1.0\linewidth]{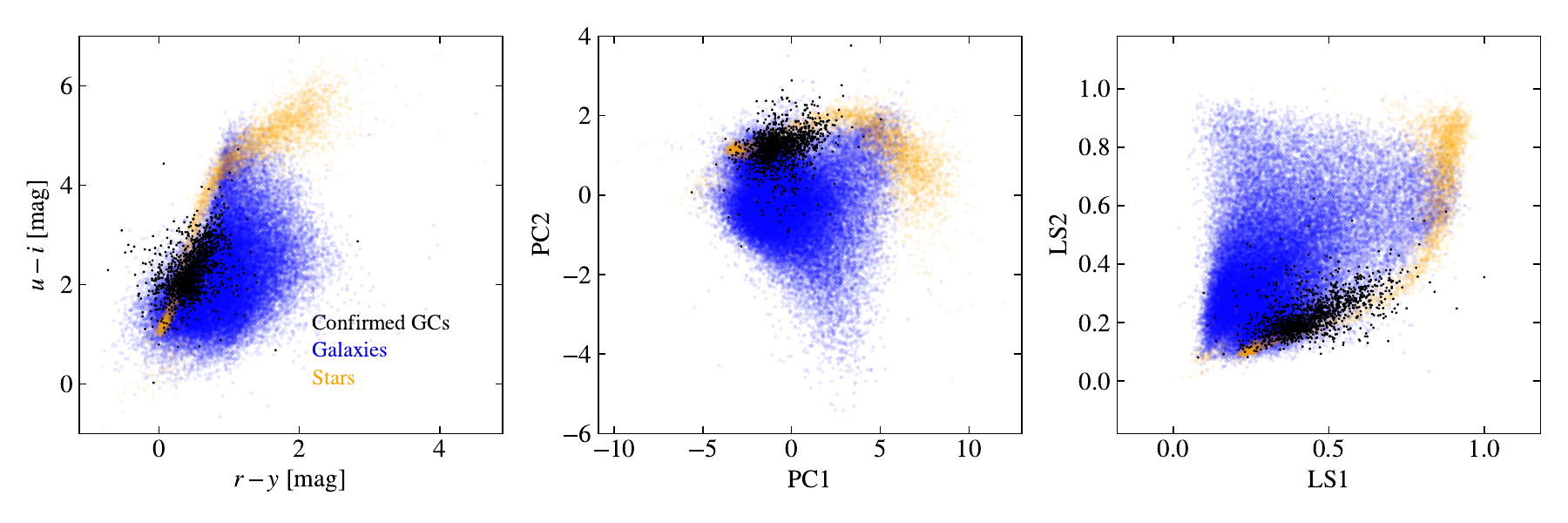}
    \caption{Projections in the color space, PC space, and the non-linear AE latent space of our LSST-like filtered and labeled photometric catalog. The plots show that there are no qualitative differences in the distribution of the points in these spaces.}
    \label{fig:projections}
\end{figure}

Figure~\ref{fig:projections} shows our dataset projected into the three different spaces of interest: 
a usual color-color diagram, the PC1-PC2 diagram, and the 2D latent space of one of our non-linear AEs (LS1-LS2 diagram). 
The plots show no qualitative differences in the distribution of points in these spaces, which leads us to suspect in advance that transforming the colors will not facilitate the correct identification of GCs in this dataset. The role of the classification models is to allow for a quantitative evaluation of this suspicion.

\subsection{Architecture choices and model training}
The specific architecture of the neural networks in question (AEs and MLPCs) depends on the dimension of the input data.
Regarding the architecture of the two AEs, their input layers have 15 neurons, associated with the 15 input colors.
The 15 dimensions of input are to be compressed into 4 in one of the AE and 2 in the other.
Hence, the layers to represent the latent spaces have 4 and 2 neurons, respectively. 
The output layers in both networks have, again, 15 neurons as the AEs are designed to attempt to reconstruct the input.
The encoders and decoders are chosen to be composed of only 1 encoding and 1 decoding layer, with 7 neurons each for the case of the 4D latent space and with 5 neurons each for the 2D one. We choose the sigmoid function (non-linear) as the activation of the encoder and a linear activation function for the decoder.
The mean squared error loss is used to train the two AEs.

With respect to the MLPCs, their input layers have 15, 4, or 2 neurons (the 15 colors, 15 PCs or fewer, 4 or 2 LSCs), 
and the output layers have 3 neurons (the problem has 3 classes). 
We decided to use 2 hidden layers, with 10 and 5 neurons, for the cases with 15 input features,
1 hidden layer with 8 neurons for the case of 4 input features, and 1 hidden layer with 4 neurons for the case of 2 input features.
For all layers, except the output one, we use the ReLU activation function. For the output layer, we use the softmax function.
The sparse categorical cross-entropy loss is used to train the MLPCs.

To properly handle class imbalance, all RFCs and MLPCs were trained using 5-fold stratified cross-validation (CV) resampling, a technique used to train and evaluate models several times, each using a separate part of the training set, while maintaining an adequate proportion of class labels as the original dataset. The RFCs were also subjected to hyperparameter tuning via randomized search, allowing the number of trees to vary, as well as the maximum depth of the tree, the minimum number of samples required to split an internal node, the minimum number of samples required to be at a leaf node, and the number of features to consider when looking for the best split. For a detailed description of the inner workings of random forest algorithms, see \cite{biauScornett2015randomforest}.
We also tested, as a pre-processing step, under-sampling the majority classes (the contaminants), over-sampling the minority classes (the confirmed GCs), and combinations of both; that is, respectively excluding and/or imputing data points to mitigate potential biases due to class imbalance when training the models. These procedures did not change the performance of the models compared to the cases in which the number of members of each class was not altered, but stratified CV was used. Hence, we decided not to use under/over-sampling methods.

In the end, the best performing models in terms of the F1-score for the GC class were evaluated on the test subset (20\% of the catalog) of each run.

\section{Results}\label{sec:results}
We ran PCA and AEs over our LSST-like photometric catalog to transform the colors into PCs and LSCs and assess how distinguishable extragalactic, point-like GCs are from the background galaxies and foreground stars using two classification models and eight input sets.
Table~\ref{tab:results} displays, for the GC class, the performance metrics (precision, recall, F1-score) of the best models for each input type. 
\begin{table}[h]
    \centering
    \begin{tabular}{lcc}
               &        RFC         &        MLPC        \\
    \hline
    15 Colors  & (0.69, 0.29, 0.41) & (0.21, 0.88, 0.34) \\
    4 Colors   & (0.55, 0.27, 0.36) & (0.20, 0.92, 0.33) \\
    2 Colors   & (0.39, 0.22, 0.28) & (0.18, 0.91, 0.30) \\
    15 PCs     & (0.71, 0.29, 0.42) & (0.21, 0.91, 0.35) \\
    4 PCs      & (0.69, 0.29, 0.41) & (0.18, 0.90, 0.30) \\
    2 PCs      & (0.35, 0.17, 0.23) & (0.17, 0.86, 0.28) \\
    4 LSCs     & (0.66, 0.32, 0.43) & (0.21, 0.90, 0.34) \\
    2 LSCs     & (0.32, 0.13, 0.19) & (0.16, 0.87, 0.27) \\
    \hline
    \end{tabular}
    \caption{Results for the classification of GCs: performance metric triplets (precision, recall, F1-score) of the best performing models (the ones with hyperparameter values that maximized the F1-score for the GC class). The GC test set contains 288 sources.}
    \label{tab:results}
\end{table}

The first direct result that can be extracted from Table~\ref{tab:results} is that the best precision score for the GC class, 71\%, was obtained by RFC, using 15 PCs as input, with a respective 29\% recall rate (71\% of the actual GCs in the test set are not identified by the model). Given that the test set was selected while preserving the original class proportions, we can view the evaluation of each model on the test set as a GC candidate selection process. In that sense, the performance of the RFC using 15 PCs as input corresponds to a sample of candidates 29\% contaminated and only 29\% complete. At the same time, the performance of the RFCs that received 15 colors, 15 PCs, and 4 PCs as input is almost identical, with a 2\% difference in precision to make the 15 PCs case stand out, which is hardly statistically significant. From the perspective of the MLPCs, with more homogeneous performance scores across the various inputs, recall rates of $\sim 90\%$ are obtained, although associated with even lower precision and, in some cases, similar F1-scores compared to the other models with the same inputs. That is, with MLPCs, despite contamination rates of $\sim 80\%$, the samples of GC candidates would contain $\sim 90\%$ of the actual GCs. The performance metrics for the classification of the background galaxies and foreground stars (the two other classes) are presented in Appendix~\ref{app:extraResults} as supplementary results. 

\begin{figure}
    \centering
    \includegraphics[width=1.0\linewidth]{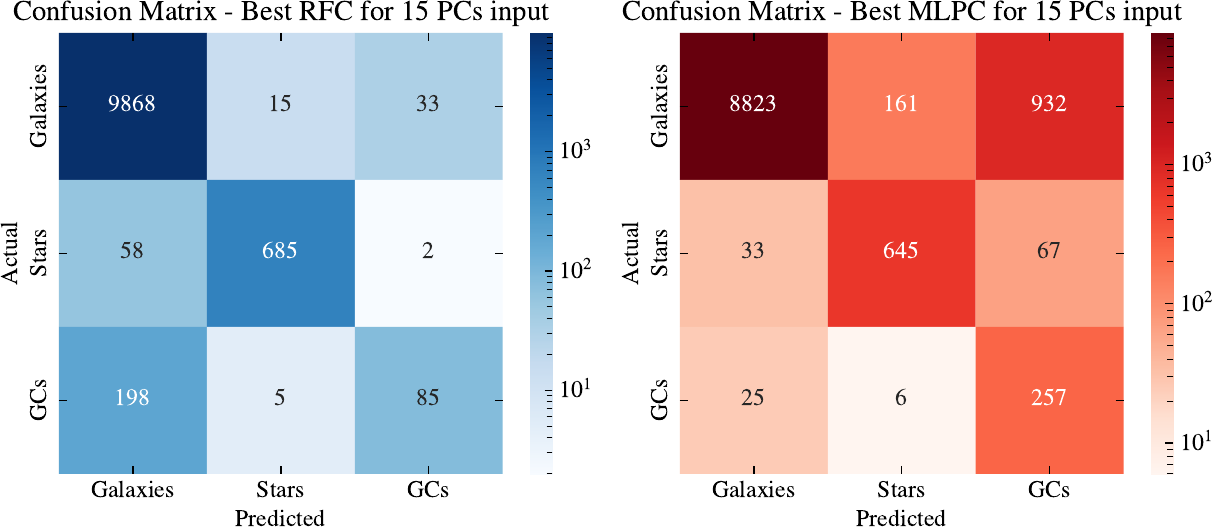}
    \caption{Confusion matrices of the RFC and MLPC that received the set of all 15 PCs as input. These are also the best performing models in terms of both precision and F1-score for the GC class, although they behave almost identically to the ones that received 15 colors as input.}
    \label{fig:confusionMatrices}
\end{figure}

Figure~\ref{fig:confusionMatrices} presents the confusion matrices (CMs) for the RFC and MLPC that receive 15 PCs as input (fourth row of Table~\ref{tab:results}); the absolute numbers in the matrices allow a more direct and detailed assessment of the precision and recall scores discussed above. This RFC misclassifies 71\% of the 288 actual GCs in the test sample, of which 69\% (198) are labeled galaxies and 2\% (5) stars (bottom row of blue CM). Although it correctly identifies 29\% of the actual GCs (85), it also misclassifies 33 galaxies and 2 stars as GCs, resulting in 35 contaminants and an overall precision of 71\% (right column of blue CM). Again, this precision score is practically the same as the one achieved by the RFC with the 15 colors input. 
From the perspective of the MLPC, in comparison with the corresponding RFC, galaxies are less confused with actual GCs(bottom row of red CM), but actual galaxies and stars are more confused with GCs (right column of red CM). Quantitatively, if the right columns of the two CMs are proportionally compared, the RFC is demonstrated to be more accurate when selecting GCs: MLPC misclassifies about 3 times more galaxies and stars as GCs than the RFC does, despite correctly identifying 89\% of the actual GCs; hence its lower F1-score of 35\%.
The CMs associated with all other runs are available in Appendix \ref{app:extraResults}.

In terms of the F1-scores for the GC class, the best performing RFCs are systematically superior to their MLPC counterparts. This, together with the above description of the CMs, suggests a limitation inherent to the dataset. More complex models may not improve the performance on the same dataset. Therefore, a concrete limit to the contribution of the color space to the selection of point-like extragalactic GC candidates is obtained from our LSST-like catalog: a very incomplete sample of candidates (recall $< 30\%$) may yield a minimum contamination rate of $\sim 30\%$, such that increasing completeness may lead to higher contamination.

It is also possible to demonstrate that PCA has some compression efficacy for this dataset; this is supported by the fact that the RFCs with 15 colors and 15 PCs inputs show the same performance scores as those from the RFC with 4 PCs input, while the RFC with 4 colors input reported a substantially lower precision score for the GC class. In contrast, the RFC with the 2 colors input yields a similar, or slightly superior, performance compared to the 2 PCs input RFC. Furthermore, the RFC that uses 4 LSCs as input also outperforms the one that uses 4 colors, but it is equivalent, or slightly inferior, to the one that uses 4 PCs as input.
Notably, all 2D inputs yield substantially lower precision scores compared to their higher-dimensional counterparts, including the cases of PCs and LSCs, which encode information from more than two colors. This specific outcome highlights the limitation of 2D color-color diagrams, and other 2D projections, to select GC candidates compared to the use of higher-dimensional projections.
Finally, the fact that the models that received LSCs do not outperform those associated with PCs indicates that there is no non-linear relation within the color space of this dataset that could be used to reduce contamination. This is quantitative evidence in accordance with the suspicion raised by visual inspection of the 2D projects in Figure~\ref{fig:projections}.

\section{Discussions and conclusions}
\label{sec:conclusions}

We have assembled an LSST-like (FDS + DES) photometric catalog of the central region of the Fornax Cluster and prepared labels for confirmed GCs from the literature, and background galaxies and foreground stars selected from DES DR2 data. Our goal was to understand to what extent the color space of this catalog enables us to correctly identify extragalactic, point-like GCs among contaminants. Using RFCs, we show that projecting the catalog colors onto their principal components allow for dimensionality reduction without compromising the precision of GC identification. Namely, the minimum contamination rate of $\sim 30\%$ is unchanged regardless of using all 15 available colors or only the first 4 PCs of these colors as input to RFCs. Nevertheless, such a contamination level is achieved at the expense of highly incomplete GC candidate selection. MLPCs do not yield improved performance, indicating an intrinsic limitation of the data.
It was also possible to show the use of LSCs from non-linear AEs yielded equivalent or less accurate results when compared to the use of PCs and colors.

This leads us to encourage the use of PCs of colors instead of colors themselves when selecting extragalactic GC candidates, especially in scenarios where many photometric bands are available. For instance, the ground-based Javalambre Physics of the Accelerating Universe Astrophysical Survey (J-PAS) uses a set of broad, intermediate, and narrow band filters, 59 in total, producing SEDs with more encoded information about the nature of the objects compared to $ugrizY$ SEDs; the color space of a J-PAS photometric catalog is composed of $\sim 1000$ dimensions. A pipeline to extract samples of extragalactic GCs from such a high-dimensional dataset could use PCA and RFCs as a solid starting point, although it is expected that more complex models could indeed be useful in this case.

The limited ability of $ugrizY$ colors to discriminate between stars, galaxies, and GCs is perhaps not too surprising. In general, the ultraviolet-to-near-infrared emission of galaxies and star clusters is dominated by starlight. Galaxy light can also include emission from active galactic nuclei and both absorption and emission from the interstellar medium, and is, of course, affected by redshifting. The colors of the simple stellar populations of GCs change with age as different stellar evolutionary stages dominate, but galaxies also contain stars with a range of ages. That is, completely distinguishing between the simpler star formation histories of GCs and the more complex ones of galaxies is not possible with $ugrizY$ photometry alone. 
In fact, similar data-driven investigations on distinguishing extragalactic GCs from contaminants have found that morphometric quantities may hold the most discriminatory information, except for when attempting to separate the faintest/smallest GCs from foreground stars. For instance, \cite{mohammadi2022} also employ a supervised learning setup to a set of confirmed GCs and ultra compact dwarf galaxies, background galaxies, and foreground stars in the Fornax Cluster, but using the colors and FWHMs from the $ugriJK_s$ (FDS + VISTA) filter-set. They find that the optical FWHMs were more important to separate background galaxies from GCs than any other color apart from $g-i$ and $g-r$ (which have the highest $S/N$), although using a more restricted sample of bright GCs compared to this work. 
A similar effect is observed in \cite{barbisan2022}, which uses $ugriz$ colors, magnitudes, and flux radii data of M87 from the Next Generation
Virgo Cluster Survey to construct a model to select GC candidates, again, in a supervised setup. They found optical flux radii to be more informative than colors, however with a very restricted sample of 90 extended background galaxies, together with 1160 bright spectroscopically confirmed GCs and 2346 foreground stars.

Therefore, additional steps to continue reducing contamination in samples of extragalactic GC candidates for multi-band surveys like LSST must rely on complementary, more informative data to augment the color space before attempting to leverage more complex models. Possibilities include: most importantly, morphometric properties, which are known to be very effective discriminators when sufficient spatial resolution is achieved, as in spaced-based facilities such as HST, Euclid, and the upcoming Roman Space Telescope \citep{peng2006, voggel2025, larsen2025, howell2025}; near-IR, as demonstrated by \cite{munoz2014} and \cite{cantiello2018} to be very useful; astrometric parallax from Gaia \citep{voggel2020, chies-santos2022}; and careful UV contribution (e.g. \citealt{pacheco2025}).
Specifically regarding the scenario in which space-based imaging in multiple filters is available, the use of convolutional neural networks and methods alike to identify very promising extragalactic GC candidates from the images themselves, or from the image cutouts of photometric candidates, becomes not only feasible, but perhaps desirable, especially for automated selection over large areas of the sky \citep{dold2022}.

With that in mind, we highlight the importance of collective work to fusion the scientific potentials of different facilities/surveys/collaborations and thus foster the perspectives of extragalactic GC science. 
For example, the first major effort to perform joint analysis involving Roman and Rubin/LSST data was ``OpenUniverse2024'' \citep{openUniverse2024}, which produced 4 million simulated individual images covering two overlapping areas: an approximate 70 degree field to be observed by both the LSST Wide-Fast-Deep Survey and the Roman High-Latitude Wide-Area Survey; and, the LSST ELAIS-S1 Deep-Drilling Field which is also to be observed by the Roman High-Latitude Time-Domain Survey. A large collaboration among NASA, the NSF and the DOE was involved in the ``OpenUniverse2024'' Project (which specifically included the NASA OpenUniverse team, the LSST Dark Energy Science Collaboration, the Roman High-Latitude Imaging Survey Project Infrastructure Team, and the Roman Supernova Project Infrastructure Team, as well as several other scientists). Another major joint analysis effort is underway by a Roman Wide Field Science Team. They will develop the Scarlet2 software package (a multi-band, multi-resolution astronomical source modeling framework to perform joint detection, deblending/modeling, and measurement for Roman and Rubin data. Catalog data products from this work are expected to be released in 2027 and 2028.

\begin{acknowledgments}
The authors thank the reviewer of this paper, whose comments and suggestions greatly enriched the quality of this work.
NSS acknowledges support from \textit{Laboratório Interinstitucional de e-Astronomia} (LIneA, Brazil), the Brazilian agencies \textit{Conselho Nacional de Desenvolvimento Científico e Tecnológico} (CNPq), \textit{Fundação Araucária}, and \textit{Fundação de Amparo à Pesquisa do Estado do Rio Grande do Sul} (FAPERGS). 
ACS acknowledges support from FAPERGS (grants 23/2551-0001832-2 and 24/2551-0001548-5), CNPq (grants 314301/2021-6, 312940/2025-4, 445231/2024-6, and 404233/2024-4), and CAPES (grant 88887.004427/2024-00). 
RSS acknowledges support from CNPq (grants 446508/2024-1, 315026/2025-1).
MC acknowledges support from ASI–INAF grant no. 2024-10-HH.0 (WP8420), the ESO Scientific Visitor Programme, and INAF GO-grant no. 12/2024 (P.I. M. Cantiello). TS acknowledges funding from the CNES postdoctoral fellowship programme.
JPC acknowledges support from Consejo Nacional de Investigaciones Científicas y Técnicas de la República Argentina, Agencia Nacional de Promoción Científica y Tecnológica, and Universidad Nacional de La Plata (Argentina).
\end{acknowledgments}

\begin{contribution}
Conceptualization, ACS, RSS, KD, NSS, JPC, CB; methodology, RSS, ACS, NSS; data curation, NSS; software, NSS, RSS; formal analysis, NSS; visualization, NSS; resources, ACS, CB; supervision, ACS, RSS, JPC, CB, TP, KD, MC, RM; writing—original draft preparation, NSS; writing—review and editing, NSS, ACS, KD, KR, PB, MC, JS, AIE, JPC, TS, TP, PSL, PF, RM, YOB, JG, NP; project administration, ACS, KD, NSS; funding acquisition, ACS, JPC, RM.
\end{contribution}

%



\appendix

\section{Comparison of FDS and DES photometric data}\label{app:data}
\begin{figure}
    \centering
    \includegraphics[width=1.0\linewidth]{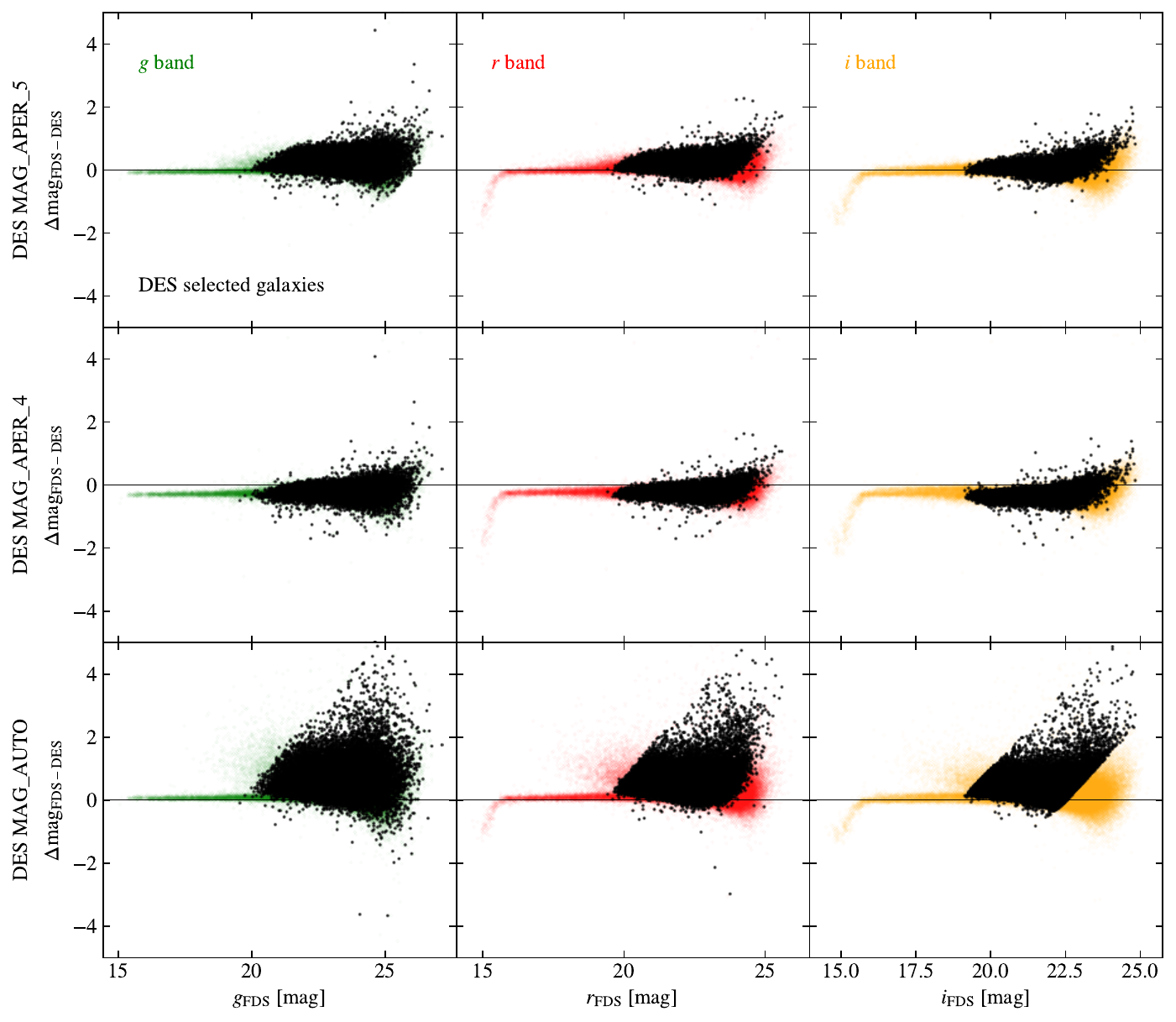}
    \caption{$\Delta\rm{mag_{FDS-DES}}$ vs $g,r,i_{\rm FDS}$: the difference in magnitude for the same source, in the same band, but in different surveys against the FDS magnitude in the same band. The first row displays the plots where DES \texttt{MAG\_APER\_5} ($2.92''$) data was used, the second DES \texttt{MAG\_APER\_4} ($1.92''$), and the third DES \texttt{MAG\_AUTO}. Black points background galaxies selected via criteria (1) as in Subsection~\ref{subsec:des}. The horizontal black line is $y=0$.}
    \label{fig:magDiffGalaxiesInBlack}
\end{figure}

\begin{figure}
    \centering
    \includegraphics[width=1.0\linewidth]{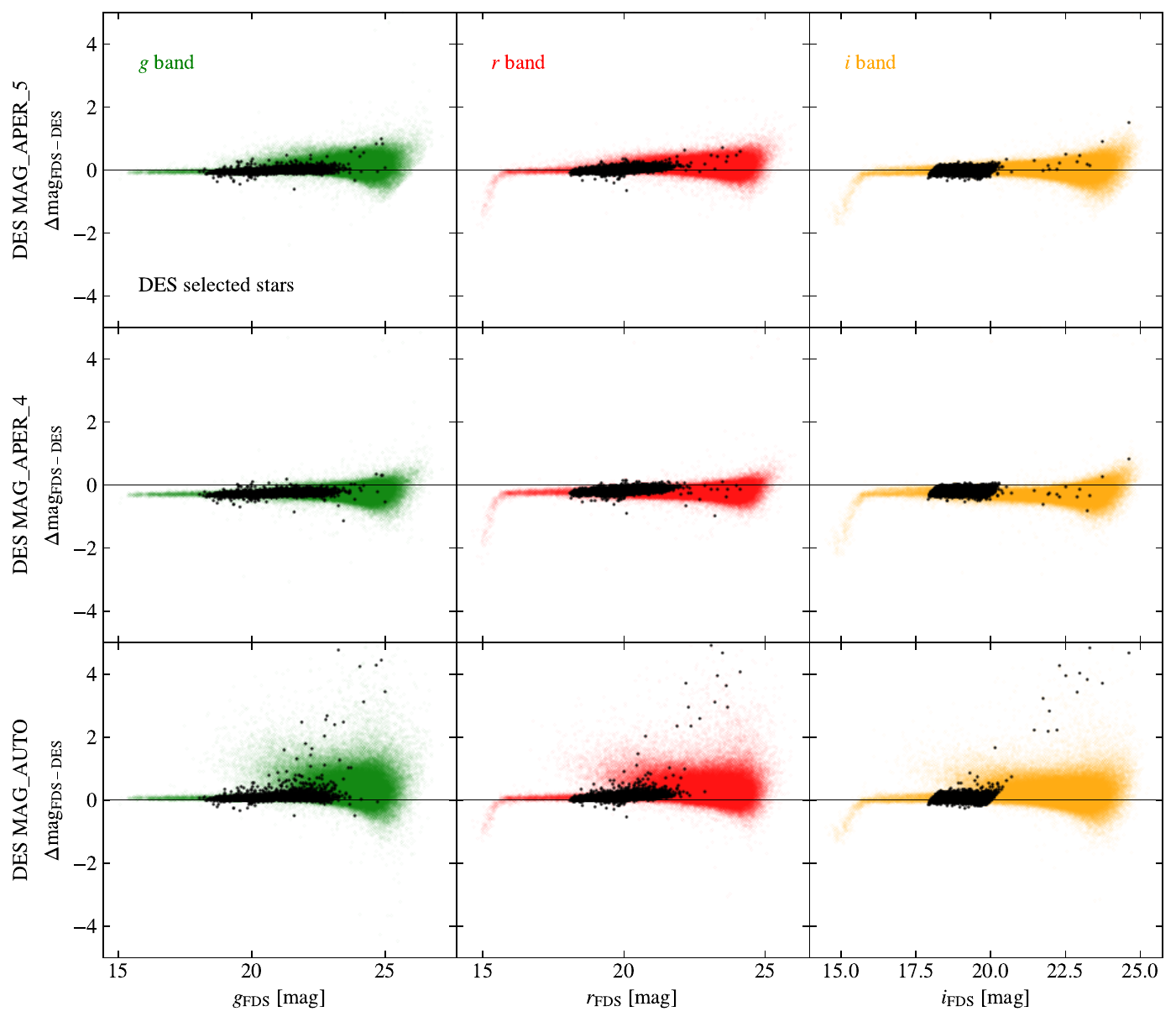}
    \caption{$\Delta\rm{mag_{FDS-DES}}$ vs $g,r,i_{FDS}$: the difference in magnitude for the same source, in the same band, but in different surveys against the FDS magnitude in the same band. The first row displays the plots where DES \texttt{MAG\_APER\_5} ($2.92''$) data was used, the second DES \texttt{MAG\_APER\_4} ($1.92''$), and the third DES \texttt{MAG\_AUTO}. Black points foreground stars selected via criteria (2) as in Subsection~\ref{subsec:des}. The horizontal black line is $y=0$.}
    \label{fig:magDiffStarsInBlack}
\end{figure}

Figures \ref{fig:magDiffGalaxiesInBlack} and \ref{fig:magDiffStarsInBlack} allow us to draw the same conclusion as obtained from the inspection of Figure~\ref{fig:magDiff}: \texttt{MAG\_APER\_5} is the circular aperture model used to perform photometry on DES images that most closely resembles FDS PSF photometry.

\section{Supplementary results}\label{app:extraResults}
\begin{figure}
    \centering
    \includegraphics[width=1.0\linewidth]{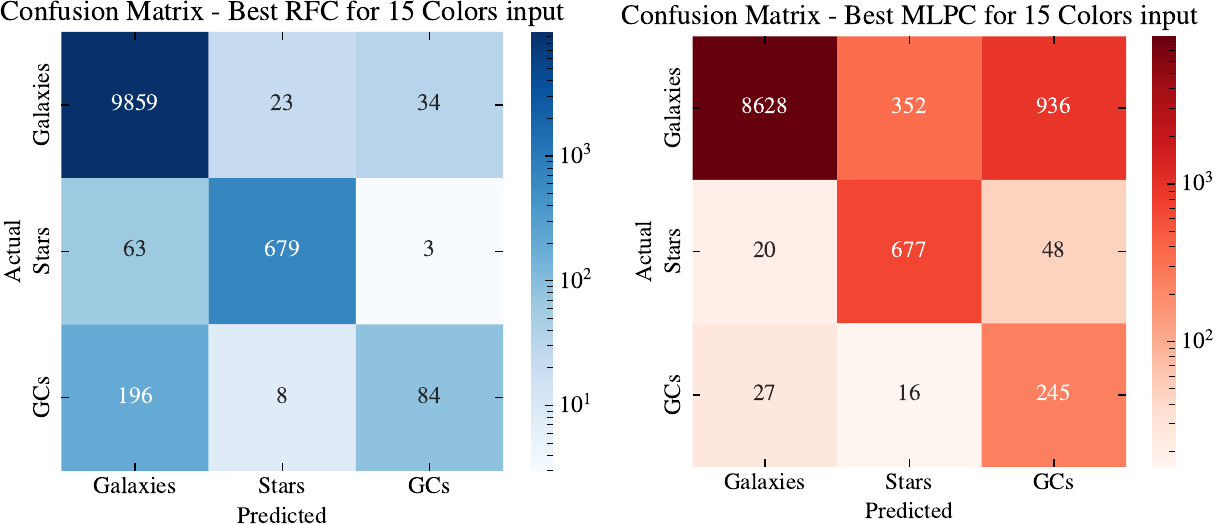}
    \caption{Confusion matrices of the RFC and MLPC that received 15 colors as input.}
    \label{fig:cm15colors}
\end{figure}

\begin{figure}
    \centering
    \includegraphics[width=1.0\linewidth]{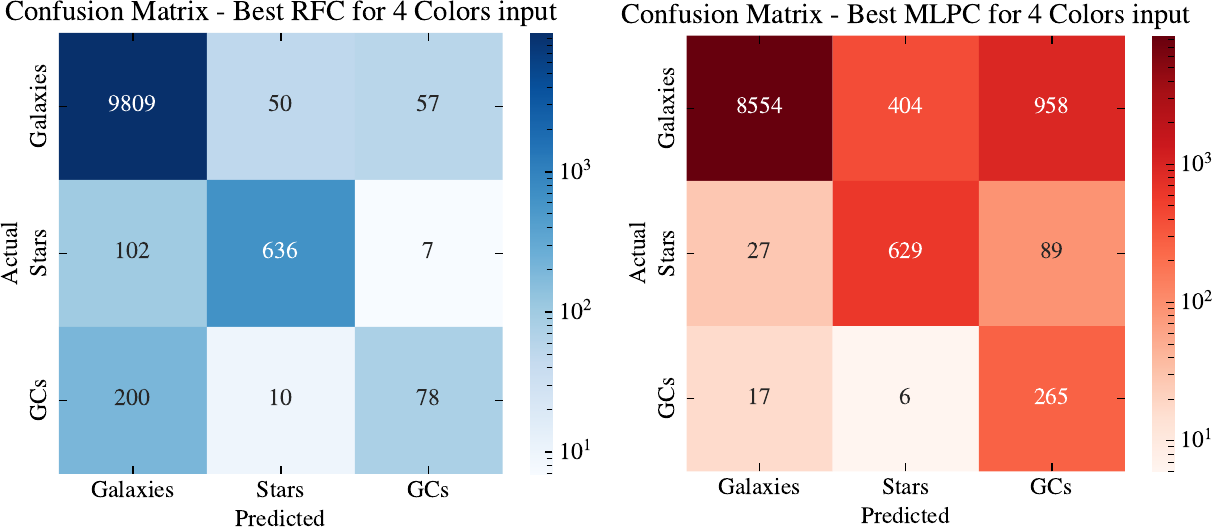}
    \caption{Confusion matrices of the RFC and MLPC that received 4 colors as input.}
    \label{fig:cm4colors}
\end{figure}

\begin{figure}
    \centering
    \includegraphics[width=1.0\linewidth]{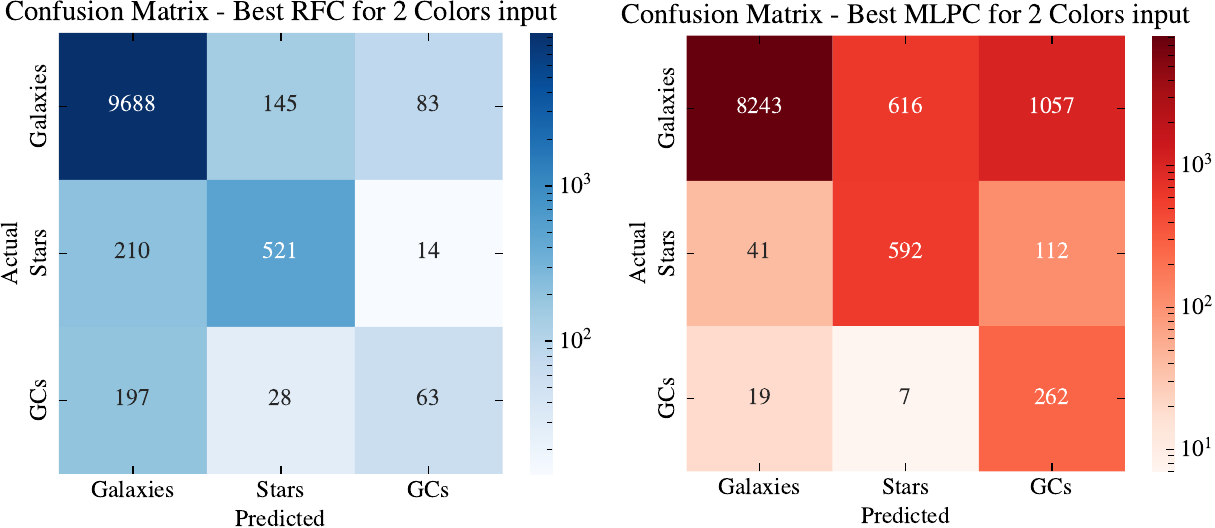}
    \caption{Confusion matrices of the RFC and MLPC that received 2 colors as input.}
    \label{fig:cm2colors}
\end{figure}

\begin{figure}
    \centering
    \includegraphics[width=1.0\linewidth]{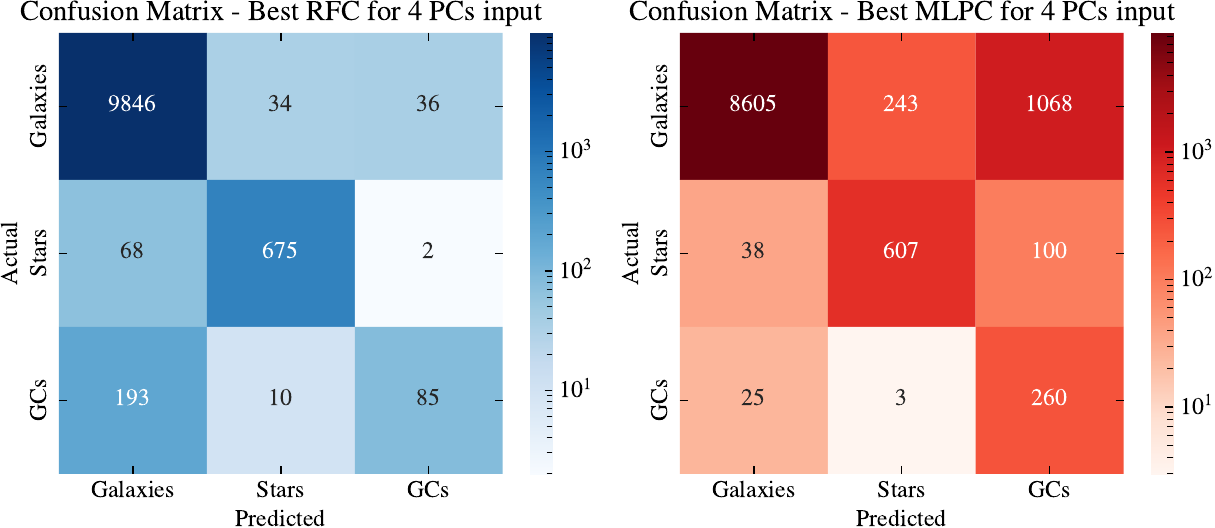}
    \caption{Confusion matrices of the RFC and MLPC that received 4 PCs as input.}
    \label{fig:cm4pcs}
\end{figure}

\begin{figure}
    \centering
    \includegraphics[width=1.0\linewidth]{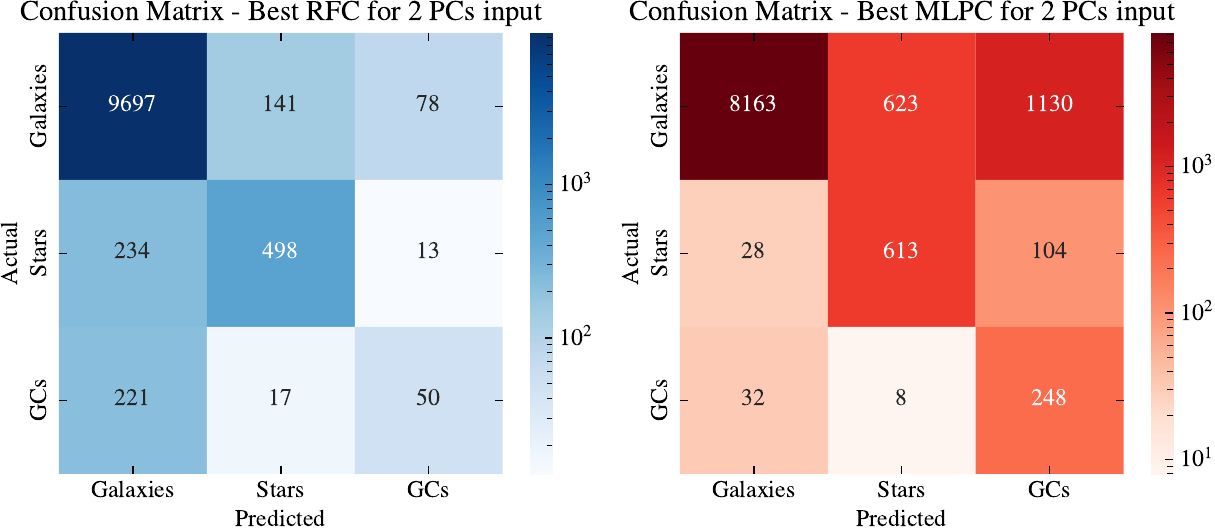}
    \caption{Confusion matrices of the RFC and MLPC that received 2 PCs as input.}
    \label{fig:cm2pcs}
\end{figure}

\begin{figure}
    \centering
    \includegraphics[width=1.0\linewidth]{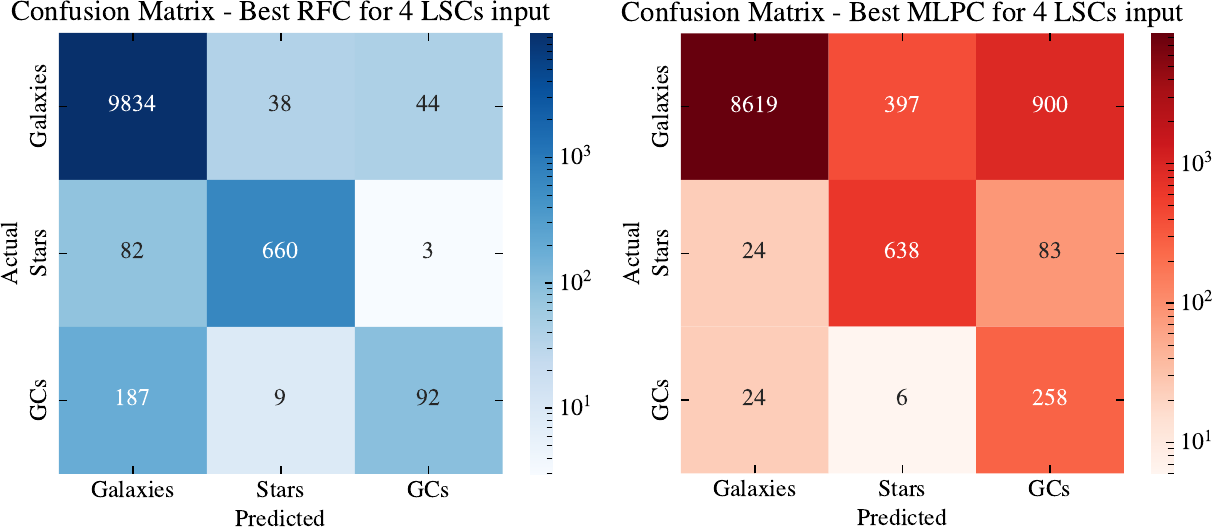}
    \caption{Confusion matrices of the RFC and MLPC that received 4 LSCs as input.}
    \label{fig:cm4lscs}
\end{figure}

\begin{figure}
    \centering
    \includegraphics[width=1.0\linewidth]{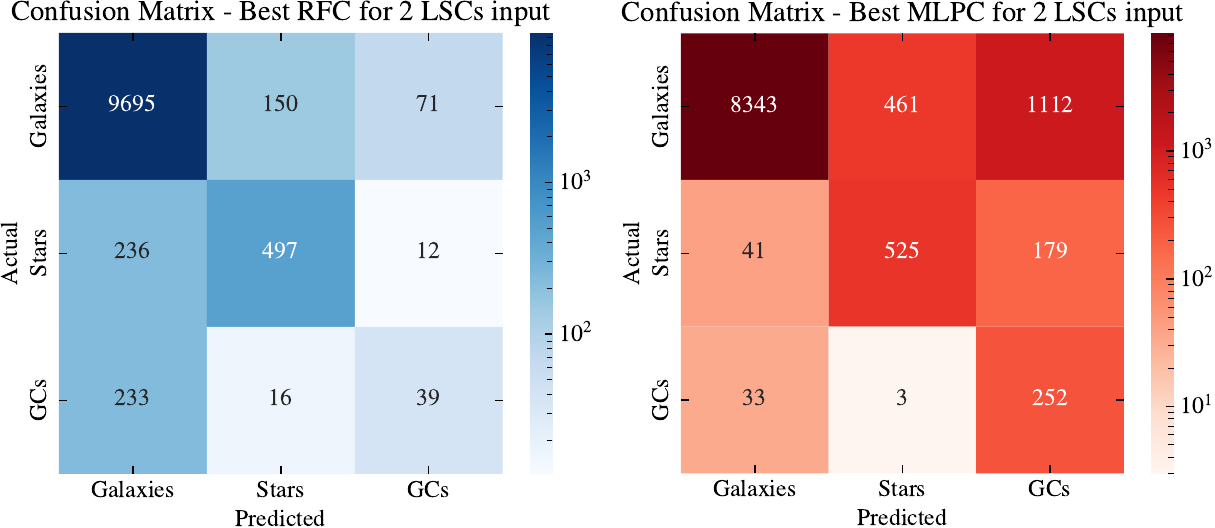}
    \caption{Confusion matrices of the RFC and MLPC that received 2 LSCs as input.}
    \label{fig:cm2lscs}
\end{figure}

In Tables \ref{tab:resultsGal} and \ref{tab:resultsStar}, we present the models' performance scores on the classification of background galaxies and foreground stars, respectively. Figures \ref{fig:cm15colors}, \ref{fig:cm4colors}, \ref{fig:cm2colors}, \ref{fig:cm4pcs}, \ref{fig:cm2pcs}, \ref{fig:cm4lscs}, and \ref{fig:cm2lscs} are the confusion matrices of all other model runs apart from those shown in Figure \ref{fig:confusionMatrices}.

\begin{table}[h]
    \centering
    \begin{tabular}{lcc}
               &        RFC         &        MLPC        \\
    \hline
    15 Colors  & (0.97, 0.99, 0.98) & (0.99, 0.87, 0.93) \\
    4 Colors   & (0.97, 0.99, 0.98) & (0.99, 0.86, 0.92) \\
    2 Colors   & (0.96, 0.98, 0.97) & (0.99, 0.83, 0.90) \\
    15 PCs     & (0.97, 0.99, 0.98) & (0.99, 0.88, 0.93) \\
    4 PCs      & (0.97, 0.99, 0.98) & (0.99, 0.87, 0.93) \\
    2 PCs      & (0.95, 0.98, 0.97) & (0.99, 0.82, 0.90) \\
    4 LSCs     & (0.97, 0.99, 0.98) & (0.99, 0.87, 0.93) \\
    2 LSCs     & (0.95, 0.98, 0.97) & (0.99, 0.84, 0.91) \\
    \hline
    \end{tabular}
    \caption{Results for the classification of background galaxies: output metric triplets (precision, recall, F1-score) of the best performing models (the ones whose hyperparameter values maximized the F1-score for the galaxy class). The galaxy test set contains 9916 sources.}
    \label{tab:resultsGal}
\end{table}

\begin{table}[h]
    \centering
    \begin{tabular}{lcc}
               &        RFC         &        MLPC        \\
    \hline
    15 Colors  & (0.96, 0.91, 0.93) & (0.64, 0.89, 0.74) \\
    4 Colors   & (0.91, 0.85, 0.88) & (0.60, 0.84, 0.70) \\
    2 Colors   & (0.75, 0.70, 0.72) & (0.49, 0.79, 0.60) \\
    15 PCs     & (0.97, 0.92, 0.94) & (0.67, 0.86, 0.75) \\
    4 PCs      & (0.94, 0.91, 0.92) & (0.71, 0.81, 0.76) \\
    2 PCs      & (0.76, 0.67, 0.71) & (0.49, 0.82, 0.62) \\
    4 LSCs     & (0.93, 0.89, 0.91) & (0.61, 0.86, 0.71) \\
    2 LSCs     & (0.75, 0.67, 0.71) & (0.53, 0.70, 0.60) \\
    \hline
    \end{tabular}
    \caption{Results for the classification of foreground stars: output metric triplets (precision, recall, F1-score) of the best performing models (the ones whose hyperparameter values maximized the F1-score for the star class). The star test set contains 745 sources.}
    \label{tab:resultsStar}
\end{table}

\bibliography{references}{}
\bibliographystyle{aasjournalv7}



\end{document}